\documentclass[12pt]{article}
\usepackage{latexsym}
\usepackage{amssymb,amsfonts,amsmath}
\usepackage{graphicx} 
\usepackage{indentfirst}
\usepackage{bbm}
\usepackage{amssymb}
\usepackage{verbatim}
\usepackage{amsmath, amsthm,amssymb}
\usepackage{mathrsfs}
\usepackage{hyperref}
\usepackage{amsfonts}
\usepackage{dsfont}

\parskip=\medskipamount

\newcommand {\cA}{{\cal A}}

\newcommand {\cC}{{\cal C}}
\newcommand {\cD}{{\cal D}}

\newcommand {\cH}{{\cal H}}

\newcommand {\cL}{{\cal L}}
\newcommand {\cM}{{\cal M}}
\newcommand {\cN}{{\cal N}}

\newcommand {\cQ}{{\cal Q}}
\newcommand {\cR}{{\cal R}}
\newcommand {\cS}{{\cal S}}
\newcommand {\cT}{{\cal T}}

\newcommand {\cW}{{\cal W}}

\def\a{\alpha}

\def\b{\beta}
\def\c{\chi}
\def\d{\delta}

\def\f{\phi}
\def\g{\gamma}

\def\k{\kappa}
\def\l{\lambda}
\def\m{\mu}

\def\o{\omega}

\def\q{\theta}
\def\r{\rho}
\def\s{\sigma}

\def\x{\xi}
\def\z{\zeta}
\def\D{\Delta}

\def\J{\Psi}
\def\L{\Lambda}
\def\O{\Omega}

\def\tr{{\rm tr}}
\def\rd{{\rm d}}
\def\ri{{\rm i}}
\def\re{{\rm e}}

\newcommand{\ad}{{\dot{\alpha}}}                           
\newcommand{\ve}{\varepsilon}                            

\newcommand{\pa}{\partial}                           
\newcommand{\hf}{\frac12}

\newcommand{\vf}{\varphi}

\newcommand{\be}{\begin{equation}}
\newcommand{\ee}{\end{equation}}
\newcommand{\bea}{\begin{eqnarray}}
\newcommand{\eea}{\end{eqnarray}}
\newcommand{\non}{\nonumber}
\newcommand{\1}{{\underline{1}}}

\newcommand{\bm}[1]{\mbox{\boldmath$#1$}}

\def\double #1{#1{\hbox{\kern-2pt $#1$}}}

\newif\ifdtup

\newcommand{\bsubeq}{\begin{subequations}}
\newcommand{\esubeq}{\end{subequations}}

\numberwithin{equation}{section}

\begin{document}

\begin{titlepage}
\begin{flushright}
June, 2018\\
\end{flushright}
\vspace{5mm}

\begin{center}
{\Large \bf Topologically massive higher spin gauge theories 
}\\ 
\end{center}

\begin{center}

{\bf 
Sergei M. Kuzenko and Michael Ponds 
} \\
\vspace{5mm}

\footnotesize{
{\it Department of Physics M013, The University of Western Australia\\
35 Stirling Highway, Crawley W.A. 6009, Australia}}  
\vspace{2mm}
~\\
Email: \texttt{sergei.kuzenko@uwa.edu.au, michael.ponds@research.uwa.edu.au
}
\\
\vspace{2mm}

\end{center}

\begin{abstract}
\baselineskip=14pt

We elaborate on conformal higher-spin gauge theory 
in three-dimensional (3D) curved space. For any integer $n>2$ we introduce
a conformal  spin-$\frac{n}{2}$ gauge field $h_{(n)} =h_{\alpha_1\dots \alpha_n}$ 
(with $n$ spinor indices) of dimension $(2-n/2)$ 
and argue that it possesses a Weyl primary descendant  $C_{(n)}$ 
of dimension $(1+n/2)$. 
The latter proves to be
divergenceless and gauge invariant in any conformally flat space.
Primary fields $C_{(3)}$ and $C_{(4)}$ coincide with 
the linearised Cottino and Cotton tensors, respectively. 
Associated with $C_{(n)}$ is a Chern-Simons-type action that is both Weyl and gauge invariant in any conformally flat space.
These actions, which  
for $n=3$ and $n=4$ coincide with the linearised actions for conformal gravitino and
conformal gravity, respectively, are used to construct 
gauge-invariant models for massive higher-spin fields in Minkowski 
and anti-de Sitter space. 
In the former case, the higher-derivative equations of motion 
are shown to be equivalent to those first-order equations which describe 
the irreducible unitary massive spin-$\frac{n}{2}$ 
representations of the 3D Poincar\'e group.
Finally, we develop ${\cal N}=1$ supersymmetric extensions of the above results. 

\end{abstract}

\vfill

\vfill
\end{titlepage}

\newpage
\renewcommand{\thefootnote}{\arabic{footnote}}
\setcounter{footnote}{0}

\tableofcontents{}
\vspace{1cm}
\bigskip\hrule

\allowdisplaybreaks


\section{Introduction} 

A unique feature of three spacetime dimensions (3D) is the existence of topologically 
massive Yang-Mills and gravity theories. 
These theories  are obtained by augmenting the usual Yang-Mills action 
or the gravitational action by a  gauge-invariant topological mass term.
Such a mass term coincides with a Chern-Simons functional
in the Yang-Mills case \cite{Siegel,JT,Schonfeld,DJT1,DJT2} 
and with a Lorentz Chern-Simons term in the case of gravity \cite{DJT1,DJT2}. 
The Lorentz Chern-Simons term is required to make the gravitational 
field possess nontrivial dynamics, 
for the pure gravity action propagates no local degrees of freedom. 
The Lorentz Chern-Simons term can be interpreted 
as the action for 3D conformal gravity
 \cite{vN,HW}.\footnote{The usual Einstein-Hilbert
action for 3D gravity with a cosmological term can also be interpreted as the Chern-Simons 
action for the anti-de Sitter group \cite{AT,Witten}.}

Topologically massive gravity possesses supersymmetric extensions.
In particular, $\cN=1$ topologically massive supergravity 
was constructed in \cite{DK,Deser}. Its topological mass term is
the supersymmetric extension of the gravitational Chern-Simons term, 
which coincides with the action for $\cN=1$ conformal supergravity \cite{vN}.
Extended topologically massive supergravity will be briefly discussed in section 7.

Topologically massive $\cN=1$ supergravity, with or without a cosmological term,
may be linearised about a maximally supersymmetric solution. 
The resulting linearised actions  
for the gravitino and the gravitational field contain higher derivatives. However, 
the genuine massive states prove to obey first-order differential equations.
This paper is devoted to the description of higher-spin extensions 
of the linearised actions for topologically massive gravity and 
$\cN=1$ supergravity.
In particular,  for every (half-)integer spin $n/2$, where $n=5,6, \dots$, we present 
a gauge-invariant higher-derivative action in Minkowski space
that propagates a single massive state of helicity $+n/2$ or $-n/2$ 
on the mass shell. The action is of the form
\bea
S_{\rm massive} = S_{\rm massless} + S_{\rm CS}~.
\label{1.1}
\eea
Here $ S_{\rm massless} $ denotes the 3D massless spin-$\frac{n}{2}$ gauge action 
of the Fronsdal-Fang type \cite{Fronsdal,FF}, with no propagating degrees of freedom. 
The second term in the right-hand side of \eqref{1.1}
is  a conformal spin-$\frac{n}{2}$ gauge
action \cite{PopeTownsend,K16}
described by a Lagrangian of the schematic form 
$\cL_{\rm CS} \propto \vf_{(n)} \pa^{n-1} \vf_{(n)}$, where $\vf_{(n)}$ stands for 
the conformal spin-$\frac{n}{2}$ field. We show that $S_{\rm massive} $
propagates a single massive state described  by the equations \eqref{2.55}.
We also present  extensions of the actions introduced
to anti-de Sitter (AdS) space, as well as
their $\cN=1$ supersymmetric generalisations. 

In the case of Minkowski space, our actions \eqref{1.1}
are in fact contained, at the component level, in the massive supersymmetric 
higher-spin models proposed in \cite{KO,KT}. However, the analysis in \cite{KO,KT}
was carried out mostly in terms of superfields
so that the component actions were not studied.
All the massive higher-spin gauge models in AdS,
 which are  presented in this paper, are new. 

This paper is organised as follows. 
In section 2 we review field realisations of the irreducible massive
spin-$\frac{n}{2}$ representations ($n = 2,3 \dots$) of the 3D Poincar\'e and AdS groups.
We also review the structure of on-shell massive higher-spin superfields
for both 3D $\cN=1$ Poincar\'e and AdS supersymmetry.
In section 3 we introduce, for any integer $n\geq 2$,  
a conformal  spin-$\frac{n}{2}$ gauge field 
${\mathfrak h}_{(n)} ={\mathfrak h}_{\a_1\dots \a_n} ={\mathfrak h}_{(\a_1\dots \a_n)}$
and argue that it possesses a Weyl primary descendant  ${\mathfrak C}_{(n)}$ 
of dimension $(1+\frac{n}{2})$ with the following properties:
(i) ${\mathfrak C}_{(n)}$ is  of the schematic form $\nabla^{n-1} {\mathfrak h}_{(n)}$;  
(ii) ${\mathfrak C}_{(n)}$ is divergenceless and gauge invariant 
in an arbitrary conformally flat space. These descendants ${\mathfrak C}_{(n)}$ 
are  constructed in any conformally flat space. 
Making use of the primary fields ${\mathfrak C}_{(n)}$, we propose
Chern-Simons-type actions 
$S_{\rm{CS}}^{(n)} \propto \int \rd^3 x \, e\, {\mathfrak h}^{\a(n)} 
{\mathfrak C}_{\a(n) }$ which are Weyl and gauge invariant in any conformally flat space,
and which 
are higher-spin extensions of the linearised action for 3D conformal gravity.
These conformal higher-spin actions are then used to construct 
massive higher-spin gauge theories in AdS, described by the actions
 \eqref{3.44b} and \eqref{3.44a}. 
In section 4 we study the dynamics of the flat-space counterparts to the gauge
theories \eqref{3.44b} and \eqref{3.44a}. 

Sections 5 and 6 are devoted to supersymmetric extensions of the results
presented in sections 3 and 4.
In section 5  we introduce conformal higher-spin gauge superfields 
${\mathfrak H}_{\a(n)}$ in curved $\cN=1$  superspace. 
These conformal gauge superfields are argued to possess primary descendants 
${\mathfrak W}_{\a(n)}$ of dimension $(1+\frac{n}{2})$ that 
are locally supersymmetric extensions of the linearised higher-spin 
super-Cotton tensors \cite{K16,KT}.
For any conformally flat superspace background, 
the primary superfields ${\mathfrak W}_{\a(n)}$ are 
explicitly constructed, and are shown to be
gauge invariant and conserved. 
Making use of ${\mathfrak H}_{\a(n)}$ and ${\mathfrak W}_{\a(n)}$,
we construct a higher-spin extension of the action for linearised $\cN=1$ 
conformal gravity, ${\mathbb S}_{\rm{SCS}}^{(n)} [ {\mathfrak H}_{(n)}] $,
which is given by eq. \eqref{5.21}.
We employ  ${\mathbb S}_{\rm{SCS}}^{(n)} [ {\mathfrak H}_{(n)}] $
to construct massive higher-spin gauge actions in $\cN=1$ AdS superspace, 
given by eqs. \eqref{5.39a} and \eqref{5.39b}.
 Section 6 describes the component structure of the supersymmetric 
higher-spin theories introduced in section 5, with the analysis being restricted 
to the flat-superspace case. Concluding comments and discussion are given in 
section 7. The main body of the paper is accompanied by three  appendices.
Appendix A describes our notation and conventions. 
Appendix B reviews the Tyutin-Vasiliev action \cite{TV}.
Appendix C provides two realisations for the higher-spin Cotton tensor in Minkowski space, $C_{\a(n)}$,  as a descendant of gauge-invariant field strengths corresponding to two different higher-spin massless models.\footnote{A similar 
result in the $\cN=2$ supersymmetric case was given in \cite{KO}.}


\section{On-shell massive (super)fields}

In this section we review the structure of irreducible massive higher-spin 
(super)fields in Minkowski space and in anti-de Sitter space.

\subsection{Massive fields}

We first recall the definition of on-shell massive fields in Minkowski space.
Given a positive integer $n>1$, a massive field, $\f_{\a_1 \cdots \a_n} 
= \bar \f_{\a_1\dots \a_n} = \f_{(\a_1 \cdots \a_n)}  $,
is a real symmetric rank-$n$ spinor field
which obeys the differential conditions \cite{TV} (see also \cite{BHT})
\begin{subequations}\label{2.55}
\bea
\pa^{\b\g} \f_{\b\g\a_1 \cdots \a_{n-2}} &=&0~, \label{dif_sub} \\
\pa^\b{}_{(\a_1} \f_{\a_2 \dots \a_n)\b} &=& m \s \f_{\a_1\dots \a_n}~,
\qquad \s =\pm 1~,
\label{mass3.6}
\eea 
\end{subequations}
with $m $ being the mass of the field.
In the spinor case, $n=1$, eq.  \eqref{dif_sub} is absent, and 
 the massive field is defined to obey the Dirac equation 
\eqref{mass3.6}.
It is easy to see that  \eqref{dif_sub} and  \eqref{mass3.6}
imply the mass-shell equation
\bea
(\Box -m^2 ) \f_{\a_1 \cdots \a_n} =0~.
\label{mass-shell}
\eea
In the spinor case, $n=1$,
eq. \eqref{mass-shell} follows from the Dirac equation \eqref{mass3.6}. 
The helicity of $\f_{\a(n)} $ is
$ \l = \frac{n}{2} \s$, 
and the spin of $\f_{\a(n)} $ is $n/2$.

It should be remarked that the system of equations \eqref{dif_sub} and 
\eqref{mass-shell} is equivalent to the 3D version of 
the Fierz-Pauli field equations \cite{FP}.  The general solution to 
 \eqref{dif_sub} and \eqref{mass-shell} is a superposition of two massive states 
of helicity $+ \frac{n}{2} $ and $-\frac{n}{2} $, respectively.
Twenty years ago, Tyutin and Vasiliev \cite{TV}
constructed Lagrangian formulations for massive higher-spin fields 
that lead to the equations \eqref{dif_sub} and \eqref{mass3.6}
on the mass shell. Their actions did not possess 
gauge invariance. In the present paper, we propose gauge-invariant 
formulations for massive higher-spin fields in
Minkowski space that lead to the equations 
\eqref{dif_sub} and \eqref{mass3.6} on-shell.

In the case of AdS space, massive fields are defined to obey the following equations
\cite{DKSS,BHRST} (see also \cite{BPSS})
\begin{subequations}\label{2.33ab}
\bea
\nabla^{\b\g} \f_{\b\g\a_1 \cdots \a_{n-2}} &=&0~, \label{2.33a} \\
\nabla^\b{}_{(\a_1} \f_{\a_2 \dots \a_n)\b} &=& \m  \f_{\a_1\dots \a_n}~,
\label{2.33b}
\eea 
\end{subequations}
for some real mass parameter $\m$. Equation \eqref{2.33b} implies that 
\bea
\big( \nabla^a \nabla_a +2(n+2) \cS^2 -\m^2 \big) \f_{\a(n)} =0~,
\label{2.44}
\eea 
where the parameter $\cS$ is related to the AdS curvature via 
eq. \eqref{3.45a}. Equation \eqref{2.44} can be rewritten in terms of 
the quadratic Casimir operator of the 3D AdS group $\rm SO(2,2)$,
\bea
{\cQ}:=\nabla^a \nabla_a
-2\mathcal{S}^2M^{\g\d}M_{\g\d}, \qquad \big[\cQ,\nabla_{a}\big]=0~,
\label{2.555}
\eea
with $M_{\g\d} $ the Lorentz generators, see Appendix A.

Equations \eqref{2.33a} and \eqref{2.44} constitute the 3D AdS counterpart to 
the  Fierz-Pauli field equations. They describe a reducible representation 
of the AdS isometry group.
Gauge-invariant Lagrangian formulations for massive higher-spin fields in AdS, 
which lead to the equations  \eqref{2.33a} and \eqref{2.44} on the mass shell, 
were developed in \cite{BSZ1,BSZ2,BSZ3,BSZ4}, 
including $\cN=1$ supersymmetric extensions obtained by combining the bosonic 
and fermionic actions (on-shell supersymmetry).
The formulations given in \cite{BSZ1,BSZ2,BSZ3,BSZ4}
are based on Zinoviev's gauge-invariant approach \cite{Zinoviev} 
to describe massive higher-spin fields.
In the present paper, we propose different gauge-invariant 
formulations for massive higher-spin fields in AdS that lead to the equations
\eqref{2.33a} and \eqref{2.33b} on-shell. 


\subsection{Massive superfields}

For $n>0$, a massive superfield $T_{\a(n)}$
is defined to be a real symmetric rank-$n$ spinor,
 $T_{\a_1 \cdots \a_n} 
= \bar T_{\a_1\dots \a_n} = T_{(\a_1 \cdots \a_n)}  $,
which obeys the differential conditions \cite{KNT-M} (see also \cite{KT})
\begin{subequations}
\label{214}
\bea
D^\b T_{\b \a_1 \cdots \a_{n-1}} &=& 0 \quad \Longrightarrow \quad
\pa^{\b\g} T_{\b\g\a_1\dots \a_{n-2}} =0
~ , 
\label{214a} 
\\
-\frac{\ri}{2} D^2  T_{\a_1 \dots \a_n} &=& m \s T_{\a_1 \dots \a_n}~, 
\qquad \s =\pm 1~.
\label{214b}
\eea
\end{subequations}
Here $D^2= D^\a D_\a$, and $D_\a$ is the spinor covariant derivative 
of $\cN=1$ Minkowski superspace.
It follows from \eqref{214a} that 
\bea
-\frac{\ri}{2} D^2  T_{\a_1 \dots \a_n} =\pa^\b{}_{(\a_1} T_{\a_2 \dots \a_n)\b} ~,
\eea
and thus $T_{\a(n)}$ is an on-shell superfield, 
\bea
\pa^\b{}_{(\a_1} T_{\a_2 \dots \a_n)\b} = m \s T_{\a_1 \dots \a_n}~,
\qquad \s =\pm 1~.
\eea
It follows from \eqref{214b} that\footnote{The equations 
\eqref{214a} and \eqref{2177} 
are the $\cN=1$ supersymmetric extension
of the Fierz-Pauli equations.} 
\bea
(\Box -m^2) T_{\a(n)}=0~.
\label{2177}
\eea
For the superhelicity of $T_{\a(n)}$  we obtain 
\bea
\k = \hf\left( n +\hf \right) \s~.
\label{218}
\eea
We define the superspin of $T_{\a(n)}$ to be $n/2$. 
The massive supermultiplet $T_{\a(n)}$ contains two ordinary 
massive fields of the type \eqref{2.55}, which are
\bea
\f_{\a_1 \dots \a_n} := T_{\a_1 \dots \a_n} |_{\q=0}~, \qquad 
\f_{\a_1 \dots \a_{n+1}} := \ri^{n+1} D_{(\a_1} T_{\a_2 \dots \a_{n+1})} |_{\q=0}~.
\eea
Their helicity values are $\frac{n}{2} \s $ and $\frac{n+1}{2}  \s$, respectively.

The off-shell gauge-invariant formulations for massive higher-spin $\cN=1$ supermultiplets in Minkowski superspace, 
which lead to the equations \eqref{214a} and \eqref{214b}  on the mass shell,
were constructed in \cite{KT}.

In the case of $\cN=1$ AdS  supersymmetry, on-shell 
 massive superfields are described by the equations \cite{KNT-M}
\begin{subequations} \label{2.12ab}
\bea
\cD^\b  T_{\a_1 \cdots \a_{n-1} \b} &=& 0~,  \label{2.12a} \\
- \frac{\ri}{2} \cD^2 T_{\a_1 \cdots \a_n} &=&\m T_{\a_1 \cdots \a_n} ~,
\label{2.12b}
\eea
\end{subequations}
with $\m$ a real mass parameter and $\cD^2 =\cD^\a \cD_\a$.
Here $\cD_A= (\cD_a, \cD_\a)$ are the covariant derivatives of the $\cN=1$ AdS 
superspace, see section 5 for the details.
 It can be shown  that 
\bea
-\frac{1}{4} \cD^2 \cD^2
=  \cD^a \cD_a -2\ri {\cS} \cD^2 
+2 {\cS} \cD^{\a \b} M_{\a\b} -2 {\cS}^2 M^{\a\b} M_{\a\b}~.
\label{2.133}
\eea
This differential operator, which is the square of the operator in the left-hand side 
of \eqref{2.12b},  
can be expressed via the quadratic 
Casimir operator\footnote{It is of interest to compare \eqref{2.155} 
with the quadratic Casimir operator of the 4D $\cN=1$ AdS supergroup
(given by eq. (29) in \cite{BKS}), which plays an important role 
in the quantisation \cite{BKS} of the massless
higher-spin supermultiplets \cite{KS94} in AdS${}_4$.}
of the 3D $\cN=1$ AdS supergroup, 
\bea
{\mathbb Q} = -\frac 14 \cD^2 \cD^2 + \ri \cS \cD^2 ~, 
\qquad \big[ {\mathbb Q}, \cD_A  \big]=0~.
\label{2.155}
\eea
It is worth pointing out that the left-hand side 
of \eqref{2.12b} can be rewritten as 
\bea
-\frac{\ri}{2} \cD^2 T_{\a_1 \cdots \a_n} 
= \cD_{(\a_1}{}^\b T_{\a_2 \cdots \a_n ) \b} 
+(n + 2) \cS T_{\a_1 \cdots \a_n} \ ,
\eea
where we have made use of \eqref{2.12a}.

In this paper we propose off-shell gauge-invariant formulations for massive higher-spin 
supermultiplets in $\cN=1$ AdS  superspace that  
lead to the equations \eqref{2.12a} and \eqref{2.12b}  on-shell.


\section{Conformal higher-spin fields}

The concept of conformal higher-spin field theory 
was introduced by Fradkin and Tseytlin in four dimensions \cite{FT}. 
(Super)conformal  higher-spin field theories in three dimensions 
were discussed in \cite{PopeTownsend,FL}. In this section, our starting 
points will be (i) the description of conformal higher-spin gauge fields in Minkowski space 
given in \cite{PopeTownsend,K16}; and (ii) the approach advocated in \cite{KMT}.

\subsection{Conformal gravity} 

The gravitational field may be described in terms of the 
torsion-free covariant derivatives
\bea
\nabla_a = e_a +\o_a = e_a{}^m  \pa_m +\hf \o_a{}^{bc}  M_{bc} ~, \qquad 
[\nabla_a , \nabla_b ] = \hf R_{ab}{}^{cd} M_{cd} ~.
\eea
Here $M_{bc} = -M_{cb}$ denotes the Lorentz generators, 
$ e_a{}^m $ the inverse vielbein, $e_a{}^m e_m{}^b = \d_a{}^b$,  
and $\o_a{}^{bc} $ the torsion-free Lorentz connection. Finally, 
$R_{ab}{}^{cd} $ is the Riemann curvature tensor.
In three dimensions, $R_{ab}{}^{cd} $ is determined by 
the Ricci tensor 
$R_{ab} := \eta^{cd} R_{c a db} = R_{ba}$ and the scalar curvature
$R=\eta^{ab}R_{ab}$. 

The Weyl tensor is identically zero in three dimensions, 
which means 
\bea
R_{abcd} = \eta_{ac} R_{bd} - \eta_{ad} R_{bc} - \eta_{bc} R_{ad} 
+ \eta_{bd} R_{ac}
-\hf (\eta_{ac} \eta_{bd} -\eta_{ad} \eta_{bc} ) R~.
\eea
The role of the Weyl tensor is played by the Cotton tensor $W_{abc} = - W_{bac}$, which is  defined in terms of 
the 3D Schouten tensor $P_{ab} = R_{ab} -\frac 14 \eta_{ab} R$ 
as follows 
\bea
W_{abc} := \nabla_a P_{bc} -\nabla_b P_{ac}~.
\label{Cotton33}
\eea
Spacetime is conformally flat if and only if the Cotton tensor vanishes
\cite{Eisen}
(see \cite{BKNT-M1} for a modern proof).
The algebraic properties of the Cotton tensor are 
\bea
W_{abc} +W_{bca} + W_{cab} =0~, \qquad W_{ab}{}^b =0~.
\eea
They imply that $W_{ab}:= \hf \ve_{acd} W^{cd}{}_b $ is symmetric and traceless,
\bea
W_{ba} =W_{ab}~, \qquad W^a{}_a =0~.
\eea
It is also divergenceless, 
\bea
\nabla^a W_{ab}=0~,
\eea
as a consequence of the Bianchi identity $\nabla^b R_{ab} = \hf \nabla_a R$.

The condition of vanishing torsion is invariant under 
Weyl (local scale) transformations of the form
\bea
\nabla_a \to \nabla'_a =  \re^{\s} \big( \nabla_a +\nabla^b\s M_{ba}\big)~,
\eea
with the parameter $\s(x)$ being completely arbitrary.  
In the infinitesimal case, the Weyl transformation laws of $R_{ab}$ and $R$ are
\bea
\d_\s R_{ab} = 2\s R_{ab} + \nabla_a \nabla_b \s +\eta_{ab} \Box \s~,
\qquad
\d_\s R = 2\s R + 4  \Box \s~,
\eea
where $\Box =\nabla^c \nabla_c $.
The Cotton tensor is a Weyl primary field of weight $+3$, 
\bea
\d_\s W_{ab} = 3\s W_{ab}~.
\eea

In what follows, we often convert every vector index into a pair of spinor ones using 
the well-known correspondence:
 a three-vector $V_a$ can equivalently be realised as a symmetric spinor 
$V_{\a\b} =V_{\b \a}$.
The relationship between $V_a$ and $V_{\a \b}$ is as follows:
\bea
V_{\a\b}:=(\g^a)_{\a\b}V_a=V_{\b\a}~,\qquad
V_a=-\hf(\g_a)^{\a\b}V_{\a\b}~.
\label{vector-rule}
\eea
Associated with the traceless part  of the Ricci tensor, $R_{ab} - \frac 13 \eta_{ab}R$,  and the Cotton tensor, $W_{ab}$, 
are the following completely symmetric rank-4 spinors:
\bea  
R_{\a\b\g\d} &:=&(\g^a)_{\a\b} (\g^b)_{\g\d} \big( R_{ab} - \frac 13 \eta_{ab}R \big)
= R_{(\a\b\g\d)}~, \\
 W_{\a\b\g\d} &:=&(\g^a)_{\a\b} (\g^b)_{\g\d}  W_{ab} 
= W_{(\a\b\g\d)} =\nabla^\r{}_{(\a} R_{\b\g \d) \r}
~.
\eea
The Weyl transformation of $R_{\a\b\g\d}$ is 
\bea
\d_\s R_{\a\b\g\d} = 2\s R_{\a\b\g\d} +\nabla_{(\a\b} \nabla_{\b\g)} \s~.
\eea


\subsection{Conformal gauge fields}

A real tensor field ${\mathfrak h}_{\a(n) } := {\mathfrak h}_{\a_1 \dots \a_n } 
={\mathfrak h}_{(\a_1 \dots \a_n)} $ is said to be a conformal 
spin-$\frac n2$ gauge field 
if (i)  it is Weyl primary of some weight $d_n$, 
\bea
\d_\s {\mathfrak h}_{\a(n)} = d_n \s {\mathfrak h}_{\a(n)}~;
\eea
and (ii) it is defined modulo gauge transformations of the form
\bea
\d_\z {\mathfrak h}_{\a(n) } =\nabla_{(\a_1 \a_2 } \zeta_{\a_3 \dots \a_n) }~,
\label{3.155}
\eea
with the real gauge parameter $\z_{\a(n-2)}$ being also Weyl primary.
These conditions uniquely fix the Weyl weight of ${\mathfrak h}_{\a(n) } $ to be
\bea
d_n = 2-\frac{n}{2}\label{2.12}~.
\eea

Starting  with ${\mathfrak h}_{\a(n) } $ one can construct its descendant,
${\mathfrak C}_{\a(n)}$, 
defined uniquely, modulo a normalisation, 
by the following the properties: 
\begin{enumerate}

\item
${\mathfrak C}_{\a(n)}$ is of the form $\cA {\mathfrak h}_{\a(n)}$, 
where $\cA$ is a linear differential operator involving 
the covariant derivatives, the curvature tensors $R_{\a(4)}$
and $R$ and their covariant derivatives.

\item
${\mathfrak C}_{\a(n)}$ is Weyl primary of weight $(1+n/2)$, 
\bea
\d_\s {\mathfrak C}_{\a(n)} = \big(1+\frac{n}{2} \big)  \s {\mathfrak C}_{\a(n)}~.
\label{2.13}
\eea

\item
The gauge variation of ${\mathfrak C}_{\a(n)}$ vanishes 
if the spacetime is conformally flat,
\bea
\d_\z {\mathfrak C}_{\a(n)} = O\big( W_{(4)}\big)~,
\label{3188}
\eea 
where $W_{(4)}$ is the Cotton tensor. 

\item
 ${\mathfrak C}_{\a(n)}$ is divergenceless if the spacetime is conformally flat,
\bea
\nabla^{\b\g} {\mathfrak C}_{\b\g \a(n-2)} = O\big( W_{(4)}\big)~.
\eea 
Here and in \eqref{3188}, 
$O\big( W_{(4)}\big)$ stands for contributions containing the Cotton tensor and
its covariant derivatives.
\end{enumerate}

We now consider several examples. Given a conformal spin-1 gauge field
${\mathfrak h}_{\a\b } = {\mathfrak h}_{\b\a}$, 
\bea
\d_\s {\mathfrak h}_{\a\b} = \s {\mathfrak h}_{\a\b} ~,
\eea
the required Weyl primary descendant is 
${\mathfrak C}_{\a\b} = \nabla^\g{}_{(\a} {\mathfrak h}_{\b)\g} $ 
and coincides with the gauge-invariant field strength, 
${\mathfrak C}_{ab} = \nabla_a {\mathfrak h}_b - \nabla_b {\mathfrak h}_a$,
of the one-form ${\mathfrak h}_a$.
This implies that 
${\mathfrak C}_{\a(2)}$ is conserved, 
\bea
\nabla^{\b\g} {\mathfrak C}_{\b\g}=0~.
\eea

Next consider a conformal spin-$\frac 32$ gauge field ${\mathfrak h}_{\a(3)}$
(i.e. conformal gravitino),
\bea
\d_\s {\mathfrak h}_{\a(3)} = \hf \s {\mathfrak h}_{\a(3)}~.
\eea
The required Weyl primary descendant is 
\bea
{\mathfrak C}_{\a(3)}
=\frac{3}{4}\nabla_{(\a_1}{}^{\b_1} \nabla_{\a_2}{}^{\b_2} 
{\mathfrak h}_{\a_3)\b_1 \b_2}
+\frac{1}{4}\Box {\mathfrak h}_{\a(3)}
+\frac{3}{4}R_{\b_1 \b_2(\a_1\a_2}{\mathfrak h}_{\a_3)}{}^{\b_1 \b_2}
-\frac{1}{16}R {\mathfrak h}_{\a(3)}~.
\eea
Its  gauge transformation is 
\bea
\d_{\z} {\mathfrak C}_{\a(3)}=-\frac{1}{2}W_{\a(3)\b}\zeta^{\b}~.
\eea
Computing its divergence gives
\bea
\nabla^{\b\g} {\mathfrak C}_{\b\g\a}=-\frac{1}{2}W_{\a\b(3)}{\mathfrak h}^{\b(3)}~.
\eea

Our last example is a conformal spin-2 gauge field ${\mathfrak h}_{\a(4)}$
(i.e. conformal graviton),
\bea
\d_{\sigma} {\mathfrak h}_{\a(4)}=0~.
\eea
The required Weyl primary descendant of ${\mathfrak h}_{\a(4)}$ is
\begin{align}
{\mathfrak C}_{\a(4)}&
=\frac{1}{2}\nabla_{(\a_1}{}^{\b_1}\nabla_{\a_2}{}^{\b_2}
\nabla_{\a_3}{}^{\b_3} {\mathfrak h}_{\a_4)\b(3)}
+\frac{1}{2}\Box\nabla_{(\a_1}{}^{\b_1} {\mathfrak h}_{\a_2\a_3\a_4)\b_1}
+\big(\nabla_{(\a_1}{}^{\b_1}R_{\a_2\a_3}{}^{\b_2\b_3}\big) 
{\mathfrak h}_{\a_4)\b(3)}\notag\\
&+\frac{1}{12}\big(\nabla_{(\a_1}{}^{\b_1}R\big){\mathfrak h}_{\a_2\a_3\a_4)\b_1}
-\frac{1}{12}R\nabla_{(\a_1}{}^{\b_1}{\mathfrak h}_{\a_2\a_3\a_4)\b_1}+2R^{\b_1\b_2}{}_{(\a_1\a_2}\nabla_{\a_3}{}^{\b_3}{\mathfrak h}_{\a_4)\b(3)}\notag\\
&-\frac{3}{4}R^{\b_1}{}_{\d(\a_1\a_2}\nabla^{\d\b_2} {\mathfrak h}_{\a_3\a_4)\b(2)}~.
\end{align}
Its gauge transformation is
\begin{align}
\d_{\zeta} {\mathfrak C}_{\a(4)}=&
\big(\nabla^{\g\d}W_{\g(\a_1\a_2\a_3}\big)\zeta_{\a_4)\delta}
+\frac{1}{2}\big(\nabla_{(\a_1\a_2}W_{\a_3\a_4)}{}^{\b(2)}\big)\zeta_{\b(2)}-W_{\g_1(\a_1\a_2\a_3}\nabla^{\g(2)}\zeta_{\a_4)\g_2}\notag\\
&+\frac{11}{12}W_{\a(4)}\nabla^{\b(2)}\zeta_{\b(2)}
+\frac{1}{2}W^{\b}{}_{\g(\a_1\a_2}\nabla_{\a_3}{}^{\g}\zeta_{\a_4)\b}~.
\end{align}
The divergence of ${\mathfrak C}_{\a(4)}$ may be shown to be
\begin{align}
\nabla^{\b\g} {\mathfrak C}_{\b\g\a(2)}=&
-\frac{1}{2}\big(\nabla_{\g(\a_1}W^{\g\b(3)}\big) {\mathfrak h}_{\a_2)\b(3)}
+\frac{5}{12}\big(\nabla_{\a(2)}W^{\b(4)}\big) {\mathfrak h}_{\b(4)}
+W_{\a(2)}{}^{\b(2)}  \nabla^{\g(2)} {\mathfrak h}_{\b(2)\g(2)}\notag\\
&-\frac{3}{2}W_{\g_1(\a_1}{}^{\b(2)}\nabla^{\g(2)}{\mathfrak h}_{\a_2)\g_2\b(2)}
-\frac{1}{12}W^{\b(4)}\nabla_{\a(2)}{\mathfrak h}_{\b(4)}~.
\end{align}

Suppose that the spacetime under consideration is conformally flat, 
\bea
W_{\a(4)}=0~.
\label{3.30}
\eea
Then the tensor ${\mathfrak C}_{\a(n)}$ 
is gauge invariant and conserved, 
\begin{subequations} \label{2.28}
\bea
\d_\z {\mathfrak C}_{\a(n)}&=&0~,  \label{2.28a}\\ 
\nabla^{\b \g} {\mathfrak C}_{\b\g\a(n-2)} &=&0~. \label{2.28b}
\eea
\end{subequations}
These properties and the Weyl transformation law \eqref{2.13} tell us 
that the action 
\bea
S_{\rm{CS}}^{(n)} [ {\mathfrak h}_{\a(n)}] 
=\frac{\text{i}^n}{2^{\left \lfloor{n/2}\right \rfloor +1}
} \int \rd^3 x \, e\, {\mathfrak h}^{\a(n)} 
{\mathfrak C}_{\a(n) } ~, \qquad e^{-1} = \det (e_a{}^m)
\label{2.29}
\eea
is gauge and Weyl invariant, 
\bea
\d_\z S_{\rm{CS}}^{(n)} [{\mathfrak h}_{\a(n)}] =0~, \qquad 
\d_\s S_{\rm{CS}}^{(n)} [ {\mathfrak h}_{\a(n)} ] =0~.
\eea
Here $\left \lfloor{x}\right \rfloor$ denotes the floor  function; it
coincides with the integer part of a real number $x\geq 0$.
The above action is actually Weyl invariant in an arbitrary curved space. 
Condition \eqref{3.30} is required to guarantee the gauge invariance 
of $S_{\rm{CS}}^{(n)} [ {\mathfrak h}_{\a(n)}] $ for $n>2$.

It follows from the Weyl transformation law \eqref{2.13} that 
$\nabla^{\b \g} {\mathfrak C}_{\b\g\a(n-2)}$ is a primary field,
\bea
\d_\s \big( \nabla^{\b \g} {\mathfrak C}_{\b\g\a(n-2)} \big)
= \Big(2+\frac{n}{2} \Big)  \s 
\nabla^{\b \g} {\mathfrak C}_{\b \g \a(n-2)}~.
\label{2.26}
\eea
This property means that
 the conservation equation \eqref{2.28b} is Weyl invariant.
 
 
\subsection{Higher-spin Cotton tensor in Minkowski space}

The linearised higher-spin Cotton tensor in Minkowski space
will be denoted 
$C_{\a(n)}  (h)$, while the previous notation ${\mathfrak C}_{\a(n)} ({\mathfrak h})$
will be reserved for curved spacetimes. 
For $n \geq 2$, $C_{\a(n)}  (h)$ is given by the expression
\cite{K16} 
\bea
&&C_{\a(n)}  (h)
:=\frac{1}{2^{n-1}} \sum\limits_{j=0}^{ \left \lfloor{n/2}\right \rfloor }
\binom{n}{2j+1}
\Box^{j}\pa_{(\a_{1}}{}^{\b_{1}}
\dots\pa_{\a_{n-2j -1}}{}^{\b_{n-2j -1}}h_{\a_{n-2j}\dots\a_{n})
\b_1 \dots \b_{n-2j -1} }~.~~~~~
\label{2.31}
\eea
It is a descendant of the conformal field $h_{\a(n)}$ defined modulo gauge 
transformations of the form
\bea
\d h_{\a(n)} = \pa_{(\a_1\a_2} \z_{\a_3 \dots \a_n)}~.
\label{3.32}
\eea
The field strength is invariant under these gauge transformations, 
\bea
\d_\z C_{\a(n)} =0~,
\eea
and obeys the Bianchi identity
\bea
\pa^{\b\g} C_{\b\g \a_1 \dots \a_{n-2}} =0~.
\label{3.334}
\eea
The higher-spin Chern-Simons action 
 \bea
 S_{\rm CS}^{(n)} [h_{(n)}] =
 \frac{\text{i}^n}{2^{\left \lfloor{n/2}\right \rfloor +1}}   
\int \rd^3 x\, h^{\a(n)} C_{\a(n)} (h)
\label{3.39}
\eea
is conformal and  invariant under \eqref{3.32}. 

 In the case of even rank, $n=2s$, with $s=1,2,\dots$, the field strength \eqref{2.31} 
 can be shown to coincide with the bosonic higher-spin Cotton tensor 
 given originally by Pope and Townsend 
 \cite{PopeTownsend}. It reduces to the linearised Cotton tensor for $n=4$, 
 and to the Maxwell field strength for  $n=2$.
 The fermionic case, $n=2s+1$, with $s=2,\dots$, was not considered in 
 \cite{PopeTownsend}. It was presented for the first time in \cite{K16}.
 
It should be pointed out that the conformal spin-3 case, $n=6$, 
was studied for the first time in \cite{DD}.
 The spin-3/2 case, $n=3$,  
was considered in \cite{ABdeRST}. 
The field strength $C_{\a(3)}$ is the linearised version of the
Cottino vector spinor \cite{DK,GPS}.

The normalisation of  $C_{\a(n)}  (h)$ defined by \eqref{2.31}
can be explained as follows. The gauge freedom \eqref{3.32} allows us to impose 
a gauge condition 
\bea
\pa^{\b\g} h_{\b\g\a(n-2)}=0~.
\eea
Under this gauge condition, the field strength \eqref{2.31}, with $s = 1,2,\dots $,
takes the form 
\begin{subequations}
\bea
C_{\alpha(2s)}&=&\Box^{s-1}\partial^{\beta}{}_{(\a_1}h_{\a_2\dots\a_{2s})\beta}
=\Box^{s-1}\partial^{\beta}{}_{\a_1}h_{\a_2\dots\a_{2s}\beta}
~, 
\label{3.41a}
\\
C_{\alpha(2s+1)}&=&\Box^s h_{\alpha(2s+1)}~, 
\label{3.41b}
\eea
\end{subequations}
as a consequence of the identity
\bea
 \sum\limits_{j=0}^{ \left \lfloor{n/2}\right \rfloor }\binom{n}{2j+1} = 2^{n-1}~.
 \eea

The field strength \eqref{2.31} proves to be the general solution to the conservation equation
\eqref{3.334}. This result has recently been proved in \cite{HHL} in the bosonic case, 
$n=2s$,
and the proof given is quite nontrivial (see also \cite{BBB}).
An  alternative proof, 
which is valid for arbitrary integer $n>1$ and is  based on supersymmetry considerations, was given in \cite{K16}.

 
\subsection{Higher-spin Cotton tensor in conformally flat spaces}

Now we are in a position to construct ${\mathfrak C}_{\a(n)}$ in 
a curved conformally flat spacetime $\cM^3$.
Locally, the covariant derivatives $\nabla_a$ of $\cM^3$
are related to the flat-space ones by 
\bea
\nabla_a
&=&
\re^\s \big( \pa_a +  \pa^b \s M_{ba} \big)
~,
\eea
for some scale factor $\s$. The linearised higher-spin 
Cotton tensor ${\mathfrak C}_{\a(n)}$ in $\cM^3$
is related to the flat-space one, eq. \eqref{2.31}, 
by the rule
\bea
{\mathfrak C}_{\a(n)} = \re^{(1+\frac{n}{2})\s} C_{\a(n)}~. \label{3.44c}
\eea
The higher-spin gauge  field ${\mathfrak h}_{\a(n)}$ in $\cM^3$ and its 
counterpart $h_{\a(n)} $ in Minkowski space are related to each other as
\bea
{\mathfrak h}_{\a(n)} =  \re^{(2-n/2)\s} h_{\a(n)}~.
\eea

In general, it is a difficult technical problem to express ${\mathfrak C}_{\a(n)}$ 
in terms of the covariant derivatives $\nabla_a$ and the gauge potential  
${\mathfrak h}_{\a(n)}$.
As an example, let us consider the case of AdS space, 
whose geometry is described by covariant derivatives satisfying the algebra
\bea
\big[\nabla_a,\nabla_b \big]=-4\mathcal{S}^2M_{ab}
\quad \Longleftrightarrow \quad 
\big[ {\nabla}_{ \a \b}, {\nabla}_{\g \delta} \big]=4\cS^2 \Big(\ve_{\g (\a}{M}_{\b) \delta} +\ve_{\delta (\a}{M}_{\b) \g}\Big).
\label{3.45a}
\eea
Here the parameter $\mathcal{S}$ is related to the AdS scalar curvature as $R= -24 \mathcal{S}^2$. 
The Cotton tensor \eqref{3.44c} for the cases $n=3,4,5$ and $6$ proves to be
\begin{align}
\mathfrak{C}_{\a(3)}=&\frac{1}{2^2}\bigg(3\nabla_{(\a_1}{}^{\b_1}\nabla_{\a_2}{}^{\b_2}\mathfrak{h}_{\a_3)\b_1\b_2}+\cQ 
\mathfrak{h}_{\a(3)}-9\mathcal{S}^2\mathfrak{h}_{\a(3)}\bigg)~,\\
\mathfrak{C}_{\a(4)}=&\frac{1}{2^3}\bigg(4\nabla_{(\a_1}{}^{\b_1}\nabla_{\a_2}{}^{\b_2}\nabla_{\a_3}{}^{\b_3}\mathfrak{h}_{\a_4)\b(3)}+4\cQ\nabla_{(\a_1}{}^{\b_1}\mathfrak{h}_{\a_2\a_3\a_4)\b_1}-80\mathcal{S}^2\nabla_{(\a_1}{}^{\b_1}h_{\a_2\a_3\a_4)\b_1}\bigg)~,\\
\mathfrak{C}_{\a(5)}=&\frac{1}{2^4}\bigg(5\nabla_{(\a_1}{}^{\b_1}\nabla_{\a_2}{}^{\b_2}\nabla_{\a_3}{}^{\b_3}\nabla_{\a_4}{}^{\b_4}\mathfrak{h}_{\a_5)\b(4)}
+10\cQ\nabla_{(\a_1}{}^{\b_1}\nabla_{\a_2}{}^{\b_2}\mathfrak{h}_{\a_3\a_4\a_5)\b(2)}+\cQ^2\mathfrak{h}_{\a(5)}\notag\\
&\phantom{extra}-330\mathcal{S}^2\nabla_{(\a_1}{}^{\b_1}\nabla_{\a_2}{}^{\b_2}\mathfrak{h}_{\a_3\a_4\a_5)\b(2)}-82\mathcal{S}^2\cQ\mathfrak{h}_{\a(5)}+1425\mathcal{S}^4\mathfrak{h}_{\a(5)}\bigg)~,\\
\mathfrak{C}_{\a(6)}=&\frac{1}{2^5}\bigg(6\nabla_{(\a_1}{}^{\b_1}\nabla_{\a_2}{}^{\b_2}\nabla_{\a_3}{}^{\b_3}\nabla_{\a_4}{}^{\b_4}\nabla_{\a_5}{}^{\b_5}\mathfrak{h}_{\a_6)\b(5)}+20\cQ\nabla_{(\a_1}{}^{\b_1}\nabla_{\a_2}{}^{\b_2}\nabla_{\a_3}{}^{\b_3}\mathfrak{h}_{\a_4\a_5\a_6)\b(3)}\notag\\
&\phantom{extra}+6\cQ^2\nabla_{(\a_1}{}^{\b_1}\mathfrak{h}_{\a_2\dots\a_6)\b_1}-960\mathcal{S}^2\nabla_{(\a_1}{}^{\b_1}\nabla_{\a_2}{}^{\b_2}\nabla_{\a_3}{}^{\b_3}\mathfrak{h}_{\a_4\a_5\a_6)\b(3)}\notag\\
&\phantom{extra}-704\mathcal{S}^2\cQ\nabla_{(\a_1}{}^{\b_1}\mathfrak{h}_{\a_2\dots\a_6)\b_1}+18432\mathcal{S}^4\nabla_{(\a_1}{}^{\b_1}\mathfrak{h}_{\a_2\dots\a_6)\b_1}\bigg)~,
\end{align}
where $\cQ$ is the quadratic Casimir of the 3D AdS group, $\rm SO(2,2)$,
given by eq. \eqref{2.555}.
Each of the tensors ${\mathfrak C}_{\a(n) }$ given above can be written 
as ${\mathfrak C}_{\a(n) } ({\mathfrak h}_{(n) }) 
= \cA {\mathfrak h}_{\a(n) }$, where the linear differential operator
$\cA$  is symmetric
in the sense that 
\bea
\int \rd^3 x \, e\, {\mathfrak g}^{\a(n)} 
\cA {\mathfrak h}_{\a(n) }
= \int \rd^3 x \, e\, {\mathfrak h}^{\a(n)} 
\cA {\mathfrak g}_{\a(n) }~,
\eea
for arbitrary prepotentials $ {\mathfrak g}_{\a(n)}$ 
and ${\mathfrak h}_{\a(n) }$. This means that it suffices to prove one of the two 
properties in \eqref{2.28}, and then the second property follows.


\section{Massive higher-spin actions in maximally symmetric  spaces} 

The conformal higher-spin actions in conformally flat spaces, eq. \eqref{2.29}, 
are formulated in terms of the gauge fields ${\mathfrak h}_{\a(n)}$. 
The same gauge field can be used to construct massless Fronsdal-Fang-type 
actions \cite{Fronsdal,FF,Fronsdal2,FF2}
 in maximally symmetric spaces. Such actions however, will involve
not only ${\mathfrak h}_{\a(n)}$ but also some compensators.

Here we describe these massless higher-spin gauge actions
in AdS${}_3$ and then use them to construct gauge-invariant models for  
massive higher-spin fields.

\subsection{Massive higher-spin actions in AdS space} 

There are two types of the higher-spin massless actions,
 first-order and second-order ones. 
Given an integer $n\geq 4$, the first-order model is described by real fields 
${\mathfrak h}_{\a(n)},\,{\mathfrak y}_{\a(n-2)}$ and ${\mathfrak y}_{\a(n-4)}$ 
which  are defined modulo gauge transformations of the form 
\begin{subequations}
\bea
\d_{\z} {\mathfrak h}_{\a(n)}&=&\nabla_{(\a_1\a_2}\zeta_{\a_3\dots\a_n)} ~,\\
\d_{\z} {\mathfrak y}_{\a(n-2)}&=&
\frac{1}{n}\nabla^\b{}_{(\a_1} \zeta_{\a_2\dots\a_{n-2})\b }
+\mathcal{S}\z_{\a(n-2)}~,\\
\d_{\z} {\mathfrak y}_{\a(n-4)}&=&\nabla^{\b(2)}\z_{\a(n-4)\b(2)}~.
\eea
\end{subequations}
The  Fang-Fronsdal-type gauge-invariant action, 
$S_{\rm{FF}}^{(n)} 
= S_{\rm{FF}}^{(n)} [{\mathfrak h}_{(n)},{\mathfrak y}_{(n-2)}, {\mathfrak y}_{(n-4)}]$,
is
\begin{align}
S_{\rm{FF}}^{(n)}=\frac{\text{i}^n}{2^{\lceil n/2\rceil}}&\int\text{d}^3x\, e\,
\bigg\{ {\mathfrak h}^{\a(n-1)\g}\nabla_{\g}{}^{\d}{\mathfrak h}_{\d\a(n-1)}
+2(n-2){\mathfrak y}^{\a(n-2)}\nabla^{\b(2)}{\mathfrak h}_{\a(n-2)\b(2)} \non\\
&+4(n-2) {\mathfrak y}^{\a(n-3)\g}\nabla_{\g}{}^{\d}
{\mathfrak y}_{\d\a(n-3)}
+2\frac{n(n-3)}{(n-1)}{\mathfrak y}^{\a(n-4)}\nabla^{\b(2)}{\mathfrak y}_{\a(n-4)\b(2)}
~~~~~\notag\\
&-\frac{(n-3)(n-4)}{(n-1)(n-2)}{\mathfrak y}^{\a(n-5)\g}
\nabla_{\g}{}^{\d} {\mathfrak y}_{\d\a(n-5)}
+(n-2)\mathcal{S} {\mathfrak h}^{\a(n)}{\mathfrak h}_{\a(n)}\notag\\
&-4n(n-2)\mathcal{S}{\mathfrak y}^{\a(n-2)}{\mathfrak y}_{\a(n-2)}
-\frac{n(n-3)}{(n-1)}\mathcal{S}{\mathfrak y}^{\a(n-4)}{\mathfrak y}_{\a(n-4)}\bigg\}~.
\label{33.41}
\end{align}
Here $\lceil n/2\rceil$ stands for the ceiling function, which is equal to $s $ for 
$n=2s$ and $s+1$ for $n=2s+1$, with $s\geq 0$  an integer. 

Given an integer $n\geq 4$, the second-order model is described by real fields 
${\mathfrak h}_{\a(n)}$ and ${\mathfrak y}_{\a(n-4)}$ 
defined modulo gauge transformations of the form 
\begin{subequations}
\bea
\d_{\z} {\mathfrak h}_{\a(n)}&=&\nabla_{(\a_1\a_2}\zeta_{\a_3\dots\a_n)}~,\\
\d_{\z} {\mathfrak y}_{\a(n-4)}&=&\frac{n-2}{n-1}\nabla^{\b(2)}\z_{\a(n-4)\b(2)}~.
\eea
\end{subequations}
The Fronsdal-type gauge-invariant action, 
$S_{\rm{F}}^{(n)} = S_{\rm{F}}^{(n)} [{\mathfrak h}_{(n)},{\mathfrak y}_{(n-4)}]$,
is
\begin{align}
S_{\rm{F}}^{(n)}=&\frac{\text{i}^n}{2^{\lfloor n/2 \rfloor +1}}
\int\text{d}^3x\, e \,
\bigg\{{\mathfrak h}^{\a(n)}\Box {\mathfrak h}_{\a(n)}
-\frac{n}{4}\nabla_{\g(2)}{\mathfrak h}^{\g(2)\a(n-2)}\nabla^{\b(2)}
{\mathfrak h}_{\a(n-2)\b(2)}\notag\\
&-\frac{n-3}{2}{\mathfrak y}^{\a(n-4)}\nabla^{\b(2)}\nabla^{\g(2)}
{\mathfrak h}_{\a(n-4)\b(2)\g(2)}
-n(n-6)\mathcal{S}^2{\mathfrak h}^{\a(n)}{\mathfrak h}_{\a(n)}\notag\\
&-\frac{(n-3)}{n}\bigg[2{\mathfrak y}^{\a(n-4)}\Box {\mathfrak y}_{\a(n-4)}
-2(n^2-2n+4)\mathcal{S}^2{\mathfrak y}^{\a(n-4)}{\mathfrak y}_{\a(n-4)}\notag\\
&\qquad \qquad 
+\frac{(n-4)(n-5)}{4(n-2)}\nabla_{\g(2)}{\mathfrak y}^{\g(2)\a(n-6)}
\nabla^{\b(2)}{\mathfrak y}_{\b(2)\a(n-6)}\bigg]\bigg\}~.
\label{33.43}
\end{align}

Our action \eqref{33.41} is a unique gauge-invariant extension to AdS space 
of the flat-space action given by Tyutin and Vasiliev \cite{TV}, 
see Appendix B for a review.
When $n$ is odd, $n=2s +1$, \eqref{33.41} 
is the unique gauge-invariant 3D counterpart 
to the Fang-Fronsdal action in AdS${}_4$ \cite{FF2}.\footnote{It is worth pointing out 
that the Fang-Fronsdal action for a massless spin-$(s+\hf)$ field  \cite{FF}
is also described in terms of a triplet of fermionic gauge fields, 
$\J_{\a(s+1) \ad(s)}$,  $\J_{\a(s-1) \ad(s)}$ and $\J_{\a(s-1) \ad(s-2)}$
and their conjugates, if one makes use of the two--component spinor notation,
see section 6.9 of \cite{BK}. 
More generally, there exist bosonic and fermionic higher-spin triplet 
models in higher dimensions \cite{FS,ST,FPT,Sorokin:2008tf,AAS}.
On-shell supersymmetric formulations for the generalised 
triplets in diverse dimensions have recently been given in 
\cite{Sorokin:2018djm}. 
}
When $n$ is even, $n=2s$, our action \eqref{33.43}
is the unique gauge-invariant 3D counterpart 
to the Fronsdal action in AdS${}_4$ \cite{Fronsdal2}.
The Fronsdal action \cite{Fronsdal2}
can also be generalised to $d$-dimensional AdS backgrounds
\cite{Segal,Buchbinder:2001bs}. Such an action in AdS${}_d$ is formulated in terms 
of a symmetric double-traceless field 
and it is fixed  by the condition of gauge invariance.\footnote{The dynamical equations
for massless higher-spin fields in AdS${}_d$ were  studied by Metsaev
\cite{M1,M2,M3,M4}. For alternative descriptions of massless higher-spin dynamics
in AdS${}_d$, see \cite{CF,FMM}.
}

Separately, each of the gauge-invariant actions  \eqref{2.29}, \eqref{33.41}
and \eqref{33.43} proves to
describe no propagating degrees of freedom.
We claim that the following models
\begin{subequations} \label{3.44ab}
\bea
S_{\rm massive}^{(2s+1)}
&=&\lambda S_{\rm{CS}}^{(2s+1)}[{\mathfrak h}_{(2s+1)}]
+\mu^{2s-1}S_{\rm{FF}}^{(2s+1)}
[{\mathfrak h}_{(2s+1)},{\mathfrak y}_{(2s-1)},
{\mathfrak y}_{(2s-3)}]
\label{3.44b}
\\
S_{\rm massive}^{(2s)}
&=&\lambda S_{\rm{CS}}^{(2s)}[{\mathfrak h}_{(2s)}]
+\mu^{2s-3}S_{\rm{F}}^{(2s)} [{\mathfrak h}_{(2s)},{\mathfrak y}_{(2s-4)}]
\label{3.44a}
\eea
\end{subequations}
describe irreducible massive fields in AdS${}_3$. Here the parameter $\l$
is dimensionless,  while $\m$ has dimension of mass.
Since we do not have a closed form expression for ${\mathfrak C}_{\a(n)}$ in AdS${}_3$, 
for arbitrary $n$, our analysis below will be restricted to the case of Minkowski space,
${\mathbb M}^3$.


\subsection{Massive higher-spin actions in Minkowski space:
The fermionic case} 

In this section we study the dynamics of the flat-space counterparts to the gauge
theories \eqref{3.44b} and \eqref{3.44a}. In fact, the resulting flat-space actions 
are contained at the component level in the massive supersymmetric 
higher-spin models proposed in \cite{KO,KT}. However, the analysis 
in \cite{KO,KT} was carried out mostly in terms of superfields 
so that the component actions were not studied in detail.



We first analyse the flat-space limit of the fermionic model 
\eqref{3.44b}. It is described by the action
\bea
S_{\rm massive}^{(2s+1)}
&=&\lambda S_{\rm{CS}}^{(2s+1)}[{h}_{(2s+1)}]
+\mu^{2s-1}S_{\rm{FF}}^{(2s+1)} [{h}_{(2s+1)},{y}_{(2s-1)},
{y}_{(2s-3)}]~,
\label{3.8}
\eea
where the massless sector is 
\begin{align}
S_{\rm{FF}}^{(2s+1)}=\frac{\text{i}}{2}&\Big(-\frac{1}{2}\Big)^s\int\text{d}^3x\,
\bigg\{h^{\a(2s)\g}\partial_{\g}{}^{\d}h_{\d\a(2s)}
+2(2s-1)y^{\a(2s-1)}\partial^{\b(2)}h_{\a(2s-1)\b(2)} \notag\\
&+4(2s-1)y^{\a(2s-2)\g}\partial_{\g}{}^{\d} y_{\d\a(2s-2)}
+\frac{2}{s}(2s+1)(s-1)y^{\a(2s-3)}\partial^{\b(2)}y_{\a(2s-3)\b(2)}
~~~~~\notag\\
&
-\frac{(s-1)(2s-3)}{s(2s-1)}y^{\a(2s-4)\g}\partial_{\g}{}^{\d}y_{\d\a(2s-4)}
\bigg\}~.
\label{3.6}
\end{align}
The action  \eqref{3.8} is invariant under the following gauge transformations:
\begin{subequations}
\bea
\d_{\z}h_{\a(2s+1)}&=&\partial_{(\a_1\a_2}\zeta_{\a_3\dots\a_{2s+1})}\label{3.7a} ~,\\
\d_{\z}y_{\a(2s-1)}&=&\frac{1}{2s+1}\partial^\b{}_{(\a_1} \zeta_{\a_2\dots\a_{2s-1})\b}\label{3.7b}~,\\
\d_{\z}y_{\a(2s-3)}&=&\partial^{\b(2)}\z_{\a(2s-3)\b(2)}\label{3.7c}~.
\eea
\end{subequations}
The equations of motion corresponding to the model \eqref{3.8} are
\begin{subequations}
\bea
0&=&\mu^{2s-1}\big(\partial^{\beta}{}_{(\alpha_1}h_{\a_2\dots\a_{2s+1})\beta}
-(2s-1)\partial_{(\a_1\a_2}y_{\a_3\dots\a_{2s+1})}\big)
+\lambda C_{\alpha(2s+1)}~,\label{3.9a}\\
0&=&\partial^{\beta(2)}h_{\alpha(2s-1)\beta(2)}
+4\partial^{\beta}{}_{(\a_1}y_{\a_2\dots\a_{2s-1})\beta}
-\frac{(s-1)(2s+1)}{s(2s-1)}\partial_{(\a_1\a_2}y_{\a_3\dots\a_{2s-1})}~,\label{3.9b}\\
0&=&(2s-1)\partial^{\beta(2)}y_{\alpha(2s-3)\beta(2)}
-\frac{2s-3}{2s+1}\partial^{\beta}{}_{(\a_1}y_{\a_2\dots\a_{2s-3})\beta}~.
\label{3.9c}
\eea
\end{subequations}

We now demonstrate that the model \eqref{3.8} indeed describes an 
irreducible massive spin-$(s+\frac{1}{2})$ field on the equations of motion.
The gauge transformation \eqref{3.7c} tells  us that  
$y_{\alpha(2s-3)}$ can be  completely gauged away, that is, we are able to impose 
the gauge condition
\begin{align}
y_{\alpha(2s-3)}=0~. \label{3.10}
\end{align} 
Then, the residual gauge freedom is described by $\z_{\alpha(2s-1)}$ constrained by 
\bea \label{3.11}
\partial^{\beta(2)}\z_{\alpha(2s-3)\beta(2)}=0 \quad
\implies  \quad
\partial^{\b}{}_{(\alpha_1}\z_{\alpha_2\dots\alpha_{2s-1})\b}
=\partial^{\b}{}_{\alpha_1}\z_{\alpha_2\dots\alpha_{2s-1}\b}~. 
\eea
In the gauge \eqref{3.10}, the equation of motion \eqref{3.9c} 
becomes the condition for $y_{\a(2s-1)}$ to be divergenceless, 
\bea\label{3.12}
\partial^{\beta(2)}y_{\alpha(2s-3)\beta(2)}=0 \quad
\implies \quad
\partial^{\b}{}_{(\alpha_1}y_{\alpha_2\dots\alpha_{2s-1})\b}
=\partial^{\b}{}_{\alpha_1}y_{\alpha_2\dots\alpha_{2s-1}\b}~.
\eea
Due to \eqref{3.12}, the gauge transformation \eqref{3.7b} becomes
\bea
\delta_\z y_{\alpha(2s-1)}=\frac{1}{2s+1}
\partial^{\b}{}_{\alpha_1}\z_{\alpha_2\dots\alpha_{2s+1}\b} ~.
\label{3.14}
\eea
Since $y_{\alpha(2s-1)}$ and $\z_{\alpha(2s-1)}$ have the same functional type, 
we are able 
to completely gauge away the $y_{\alpha(2s-1)}$ field,
\begin{align}
y_{\alpha(2s-1)}=0~. \label{3.18}
\end{align}
In accordance with \eqref{3.14} and \eqref{3.18}, 
the residual gauge freedom is described by the parameter $\z_{\alpha(2s-1)}$ 
constrained by 
\bea
\partial^{\b}_{~\alpha_1}\zeta_{\a_2\dots\a_{2s-1}\b}=0
\quad \implies  \quad \Box  \z_{\alpha(2s-1)}=0~.
\label{3.18.5}
\eea
In the gauge \eqref{3.18}, the equation of motion \eqref{3.9b} tells us that $h_{\a(2s+1)}$ is divergenceless, 
\begin{align}
\partial^{\beta(2)}h_{\alpha(2s-1)\beta(2)}=0 
\quad \implies  \quad
\partial^{\b}{}_{(\alpha_1}h_{\alpha_2\dots\alpha_{2s+1})\b}
=\partial^{\b}{}_{\alpha_1}h_{\alpha_2\dots\alpha_{2s+1}\b}~.
\label{3.19}
\end{align}
So far the above analysis has been identical to that given in Appendix B of \cite{KO} 
for the massless model \eqref{3.6}.

Due to \eqref{3.19},
the Cotton tensor \eqref{2.31} reduces
to the expression \eqref{3.41b}.
In the gauge \eqref{3.18}, 
the equation of motion \eqref{3.9a} becomes
\begin{align}
\mu^{2s-1}\partial^{\beta}{}_{\a_1}h_{\a_2\dots\a_{2s+1}\beta}
+\lambda\Box^s h_{\alpha(2s+1)}=0~.\label{3.21}
\end{align}
This equation has two types of solutions, massless and massive ones,
\begin{subequations}
\bea
\partial^{\b}{}_{\alpha_1}h_{\alpha_2\dots\alpha_{2s+1}\b}=0
\quad \implies \quad  \Box h_{\alpha(2s+1)}=0~;
\label{4.14a}\\
\mu^{2s-1} h_{\a (2s+1)}
+ 
\l
\Box^{s-1} \partial^{\beta}{}_{\a_1} h_{\alpha_2 \dots \a_{2s+1} \b}=0~.
\label{4.14b}
\eea
\end{subequations}
We point out that $\J_{\a_1 \dots \a_{2s+1}} :=\partial^{\beta}{}_{\a_1} h_{\alpha_2 \dots \a_{2s+1} \b}$ 
is completely symmetric and divergenceless, 
$\J_{\a_1 \dots \a_{2s+1}} = \J_{(\a_1 \dots \a_{2s+1})}$ and 
$ \pa^{\b\g} \J_{\b\g \a_1 \dots \a_{2s-1}} =0$.

Let us show that the massless solution \eqref{4.14a} is a pure gauge degree of freedom.
Since both the gauge field $h_{\alpha(2s+1)}$ and 
the residual gauge parameter $\z_{\alpha(2s-1)}$ are on-shell massless, 
 it is useful to switch to momentum space, by replacing 
$h_{\a(2s+1)} (x) \to h_{\a(2s+1)} (p)  $ and $\z_{\a(2s-1)}(x) \to\z_{\a(2s-1)}(p)$, where
the three-momentum $p^a$ is light-like, $p^{\a\b}p_{\a\b}=0$. 
For a given three-momentum, 
we can choose a frame in which the only non-zero component of 
$p^{\a\b}= (p^{11}, p^{12} =p^{21}, p^{22}) $ is $p^{22}=p_{11}$. 
Then, the conditions $p^\b{}_{\a_1} h_{\a_2 \dots \a_{2s+1} \b}(p)=0$ and 
$p^\b{}_{\a_1} \z_{\a_2 \dots \a_{2s-1} \b}(p)=0$ are equivalent to 
\bea
h_{\a(2s)2 }(p)=0~, \qquad \z_{\a(2s-2)2 }(p)=0~.
\eea
Thus the only non-zero components of $h_{\a(2s+1)} (p)$ and  $\z_{\a(2s-1)} (p)$ 
are $h_{1\dots 1}(p) $ and  $\z_{1\dots 1}(p) $. 
The residual gauge freedom, 
$\d h_{1 \dots } (p) \propto p_{11 } \z_{1 \dots 1}$, 
allows us to gauge away the field $h_{\a(2s+1) } $ completely.

Thus, it remains to analyse the general solution of the equation \eqref{4.14b}, 
which implies 
\begin{align}
\Big(\Box^{2s-1}-(m^2)^{2s-1}\Big) h_{\alpha(2s+1)}=0~,
\qquad m:=\Big|\frac{\mu }{\lambda^{1/(2s-1)}}\Big|~.
 \label{3.22}
\end{align}
This equation 
in momentum space yields
\begin{align}
\bigg(1-\bigg(\frac{-p^2}{m^2}\bigg)^{2s-1}\bigg)h_{\alpha(2s+1)}(p)=0~. \label{3.25}
\end{align}
Since the polynomial equation $z^{2s-1}-1=0$ has only one real root, 
$z=1$, 
the only  real solution to \eqref{3.25} is $p^2=-m^2$, 
from which it follows that $h_{\alpha(2s+1)}$ satisfies the ordinary  Klein-Gordon equation,
\begin{align}
\big(\Box-m^2\big)h_{\alpha(2s+1)}=0~. \label{3.26}
\end{align}
Applying \eqref{3.26} to \eqref{3.21} reveals that $h_{\alpha(2s+1)}$ satisfies the equation of motion corresponding to a massive spin $(s+\frac{1}{2})$-field with mass 
$m$ and helicity $\s (s+\frac{1}{2})$,
\begin{align}
\partial^{\beta}{}_{\alpha_1}h_{\a_2\dots\a_{2s+1}\beta}=\s m
h_{\alpha(2s+1)}~,\qquad \s:=-\text{sign}(\mu \lambda)~. 
\label{3.27}
\end{align}

Finally, for completeness let us recall  the proof of the fact that equation \eqref{3.27}
describes 
a single propagating degree of freedom. 
 The field $h_{\alpha(2s+1)}$ is on-shell with momentum satisfying 
 $p^2=-m^2$, we can therefore transform equation \eqref{3.27} into momentum space and boost into the rest frame where $p^a=(m,0,0) \implies p^1{}_{1}=p^2{}_{2}=0,~ p^1{}_{2}=-p^2{}_{1}=m$, 
\begin{align}
\text{i}h_{\alpha(2s)1}(p)-\s h_{\alpha(2s)2}(p)=0~. \label{3.28}
\end{align}
Due to the symmetry of the field $h_{\alpha(2s+1)}$, equation \eqref{3.28} states that there is only a single degree of freedom. Taking the independent field component to be $h_{11\dots 1}(p)$ allows us to express all other components 
in terms of it.

Along with the fermionic model \eqref{3.8}, which corresponds to $n=2s+1$,
we could consider a bosonic one described by the action
\bea
S_{\rm massive}^{(2s)}
&=&\lambda S_{\rm{CS}}^{(2s)}[{h}_{(2s)}]
+\mu^{2s-2}S_{\rm{FF}}^{(2s)} [{h}_{(2s)},{y}_{(2s-2)}, {y}_{(2s-4)}]~,
\label{4.25}
\eea
which corresponds to $n=2s$. Most of the above analysis would remain 
valid in this case as well. However, in place of eq. \eqref{3.25} we would have 
\begin{align}
\bigg(1-\bigg(\frac{-p^2}{m^2}\bigg)^{2s-2}\bigg)h_{\alpha(2s)}(p)=0~.
\end{align}
This equation has both physical ($p^2=-m^2$) and tachyonic ($p^2=m^2$)
solutions. Therefore, the model \eqref{4.25} is unphysical.
This may be interpreted  as a consequence of the spin-statistics theorem.


\subsection{Massive higher-spin actions in 
Minkowski space: 
The bosonic case} 

Our next goal is to analyse the flat-space limit of the bosonic model 
\eqref{3.44a}. It is described by the action
\bea
S_{\rm massive}^{(2s)}
&=&\lambda S_{\rm{CS}}^{(2s)}[{h}_{(2s)}]
+\mu^{2s-3}S_{\rm{F}}^{(2s)}
[{h}_{(2s)}, {y}_{(2s-4)}]~,
\label{3.31}
\eea
where the second term is
\begin{align}
S_{\rm{F}}^{(2s)}=&\frac{1}{2}\bigg(\frac{-1}{2}\bigg)^s\int\text{d}^3x\,
\bigg\{h^{\a(2s)}\Box h_{\a(2s)}-\frac{s}{2}\partial_{\g(2)}h^{\g(2)\a(2s-2)}\partial^{\b(2)}h_{\a(2s-2)\b(2)}\notag\\
&-\frac{(2s-3)}{2s}\bigg[s y^{\a(2s-4)}\partial^{\b(2)}\partial^{\g(2)}h_{\a(2s-4)\b(2)\g(2)}+2y^{\a(2s-4)}\Box y_{\a(2s-4)}\notag\\
&+\frac{(s-2)(2s-5)}{4(s-1)}\partial_{\g(2)}y^{\g(2)\a(2s-6)}\partial^{\b(2)}y_{\b(2)\a(2s-6)}\bigg]\bigg\}\label{3.29}~.
\end{align}
The action \eqref{3.31} is invariant under the gauge transformations
\begin{subequations}
\bea
\d_{\z}h_{\a(2s)}&=&\partial_{(\a_1\a_2}\z_{\a_3\dots\a_{2s})}~,\label{3.30a}\\
\d_{\z}y_{\a(2s-4)}&=&\frac{2s-2}{2s-1}\partial^{\b(2)}\z_{\a(2s-4)\b(2)}\label{3.30b}~.
\eea
\end{subequations}
The equations of motion corresponding to \eqref{3.31} are
\begin{subequations}
\begin{align}
&0=\mu^{2s-3}\Big(\Box h_{\alpha(2s)}+\frac{1}{2}s\partial^{\beta(2)}\partial_{(\a_1\a_2}h_{\a_3\dots\a_{2s})\beta(2)}+\notag\\
&\phantom{BBBBBBBBB}
-\frac{1}{4}(2s-3)\partial_{(\a_1\a_2}\partial_{\a_3\a_4}y_{\a_5\dots\a_{2s})}\Big)
+\lambda C_{\alpha(2s)}~,\label{3.32a}\\
&0=\partial^{\beta(2)}\partial^{\gamma(2)}h_{\alpha(2s-4)\beta(2)\gamma(2)}
+\frac{4}{s}\Box y_{\alpha(2s-4)}+\notag\\
&\phantom{How much wood can a woodchop chop}-\frac{(s-2)(2s-5)}{2s(s-1)}\partial^{\beta(2)}\partial_{(\a_1\a_2}y_{\a_3\dots\a_{2s-4})\beta(2)}~.
\label{3.32b}
\end{align}
\end{subequations}

We will now show that on-shell, the model $S_{\rm{massive}}^{(2s)}$ describes a massive 
spin-$s$ field which propagates a single degree of freedom. 
As follows from the gauge transformation \eqref{3.30b}, 
it is possible
 to completely gauge away $y_{\alpha(2s-4)}$,
\begin{align}
y_{\alpha(2s-4)}=0~.
\label{3.33}
\end{align}
Then, the residual gauge freedom is described by a parameter $\zeta_{\alpha(2s-2)}$
constrained by
\bea \label{3.34}
\partial^{\beta(2)}\zeta_{\alpha(2s-4)\beta(2)}=0 \quad
\implies \quad
\partial^{\b}{}_{(\alpha_1}\z_{\alpha_2\dots\alpha_{2s-2})\b}
=\partial^{\b}{}_{\alpha_1}\z_{\alpha_2\dots\alpha_{2s-2}\b}~.
\eea
In the gauge \eqref{3.33}, the equation of motion \eqref{3.32b} becomes 
\begin{align}
\partial^{\gamma(2)}\partial^{\beta(2)}h_{\alpha(2s-4)\beta(2)\gamma(2)}=0~.
\label{3.36}
\end{align}
According to \eqref{3.30a}, the divergence of $h_{\alpha(2s)}$ transforms as
\begin{align}
\delta_\z\big(\partial^{\beta(2)}h_{\alpha(2s-2)\beta(2)}\big)&=\partial^{\beta_1\beta_2}\partial_{(\alpha_1\alpha_2}\zeta_{\alpha_3\dots\alpha_{2s-2}\beta_1\beta_2)} 
=-\frac{2}{s}\Box \zeta_{\alpha(2s-2)} \label{3.37}
\end{align}
where we have made use of \eqref{3.34}.
 Since $\zeta_{\alpha(2s-2)}$ and $\partial^{\beta(2)}h_{\alpha(2s-2)\beta(2)}$
 have the same functional type,  it is possible  to completely gauge away 
 the divergence of $h_{\alpha(2s)}$,
\bea\label{3.38}
\partial^{\beta(2)}h_{\alpha(2s-2)\beta(2)}=0 \quad 
\implies \quad
\partial^{\b}{}_{(\alpha_1}h_{\alpha_2\dots\alpha_{2s})\b}
=\partial^{\b}{}_{\alpha_1}h_{\alpha_2\dots\alpha_{2s}\b}~.
\eea
Under the gauge conditions imposed, there still remains 
some residual gauge freedom described by a gauge parameter constrained by 
\eqref{3.34} and 
 $\Box \zeta_{\alpha(2s-2)}=0$. 
So far the above analysis has been identical to that given in Appendix B of \cite{KO} 
for the massless model \eqref{3.29}.

As a consequence of \eqref{3.38}, the Cotton tensor \eqref{2.31} 
reduces to the simple form \eqref{3.41a}.
Making use of the gauge conditions \eqref{3.33} and \eqref{3.38} 
in conjunction with eq.  \eqref{3.41a}, the equation of motion \eqref{3.32a} becomes
\begin{align}
\Big( \mu^{2s-3} \d^\b{}_{\a_1}
+\lambda\Box^{s-2}\partial^{\beta}{}_{\a_1}
\Big)
\Box h_{\a_2\dots\a_{2s}\beta}=0 ~.
\label{3.40}
\end{align}
This equation has two types of solutions, massless and massive ones,
\begin{subequations}
\bea
\Box h_{\a (2s)}&=&0 ~;  \label{4.30a}\\ 
\mu^{2s-3} h_{\a(2s)} 
+\lambda\Box^{s-2}\partial^{\beta}{}_{\a_1}
 h_{\a_2\dots\a_{2s}\beta}&=&0 ~. \label{4.30b}
\eea
\end{subequations}

Let us show that the massless solution \eqref{4.30a} is a pure gauge degree of freedom.
Since both the gauge field $h_{\alpha(2s)}$ 
and the gauge parameter  $\zeta_{\alpha(2s-2)}$ are on-shell 
massless, it is useful to switch to momentum space 
 by replacing 
$h_{\a(2s)} (x) \to h_{\a(2s)} (p)  $ and $\z_{\a(2s-2)}(x) \to\z_{\a(2s-2)}(p)$, where
the three-momentum $p^a$ is light-like, $p^{\a\b}p_{\a\b}=0$. 
As in the fermionic case studied in the previous subsection,  
we can choose a frame in which the only non-zero component of 
$p^{\a\b}= (p^{11}, p^{12} =p^{21}, p^{22}) $ is $p^{22}=p_{11}$. 
In this frame, the equations \eqref{3.34} and \eqref{3.38} are equivalent to
\begin{align}
h_{\alpha(2s-2)22}(p)=0 ~, \qquad  \zeta_{\alpha(2s-4)22}(p)=0~.\label{3.43}
\end{align}
These conditions tell us 
 that the only non-zero components in this frame are $h_{1\dots 1}(p)$, $h_{1\dots 12}(p)$ and $\zeta_{1\dots 1}(p)$, $\zeta_{1\dots 12}(p)$. However, the gauge transformation \eqref{3.30a} is equivalent to $\delta h_{1\dots 1}(p)\propto \zeta_{1\dots1}(p)$ and $\delta h_{1\dots 12}(p)\propto \zeta_{1\dots12}(p)$, allowing us to completely gauge away the $h_{\alpha(2s)}$ field. 

Let us turn to the other equation  \eqref{4.30b}, which implies 
\begin{align}
\Big(\Box^{2s-3}-(m^2)^{2s-3}\Big) h_{\alpha(2s)}=0~,
\qquad m:=\Big|\frac{\mu}{~~\lambda^{1/(2s-3)}}\Big|~. 
\label{3.44}
\end{align}
Here the mass parameter has the same form  
as in the fermionic case, eq. \eqref{3.22}.
Transforming eq. \eqref{3.44} to momentum space gives
\begin{align}
\bigg(1-\bigg(\frac{-p^2}{m^2}\bigg)^{2s-3}\bigg) h_{\alpha(2s)}(p)=0~.
\end{align}
In complete analogy with the fermionic 
case considered in the previous subsection, 
this equation has the unique real solution
$p^2=-m^2$. 

 It follows that $h_{\alpha(2s)}$ satisfies the Klein-Gordon equation, 
\begin{align}
(\Box-m^2)h_{\alpha(2s)}=0~.
\end{align}
As a consequence, 
the  equation of motion \eqref{4.30b} leads to
\begin{align}
\partial^{\beta}_{~\alpha_1}h_{\a_2\dots\alpha_{2s}\beta}=\s m h_{\alpha(2s)}~,
\qquad  \s:=-\text{sign}(\mu \lambda)~. 
\label{3.47}
\end{align}
Therefore $h_{\a(2s)}$ is an irreducible on-shell massive field 
with mass $m$ and helicity $\l =\s s$.
Equation \eqref{3.47} implies that $h_{\a(2s)}$ describes a single 
propagating degree of freedom.

%


\section{Conformal higher-spin gauge superfields}

Conformal higher-spin gauge superfields in $\cN=1$ Minkowski superspace 
were introduced in \cite{K16,KT}, as a by-product of the $\cN=2$ approach of \cite{KO}.
In this section we start by  
generalising this concept to the case of $\cN=1$ supergravity, 
building on the ideas advocated in \cite{KMT}.

\subsection{Conformal supergravity}

Consider a curved $\cN=1$ 
superspace, $\cM^{3|2}$, parametrised by local real coordinates
$z^{M}=(x^m,\q^{\mu})$, with $m=0,1,2$ and $\mu=1,2$,
of which $x^m$ are bosonic and $\q^{\mu}$ fermionic. 
We introduce a basis of one-forms
$E^A=(E^a,E^\a)$ and its dual basis $E_A=(E_a,E_\a)$, 
\bea
E^A=\rd z^ME_{M}{}^A~,
\qquad E_A = E_A{}^M  \pa_M ~,
\label{beins}
\eea
which will be referred to as the supervielbein and its inverse, respectively.
The superspace structure group is ${\rm SL}(2,{\mathbb R})$, 
the double cover of the connected Lorentz group ${\rm SO}_0(2,1)$. 
The covariant derivatives have the form:
\bea
\cD_{A}&=& (\cD_a, \cD_\a )= E_{A}+\O_A~,
\label{23cd}
\eea
where
\bea
\O_A=\hf\O_{A}{}^{bc}M_{bc}=-\O_{A}{}^b M_b=\hf\O_{A}{}^{\b\g}M_{\b\g}
\label{2.444}
\eea
is the Lorentz connection.

The covariant derivatives are characterised by the graded commutation relations 
\bea
{[}\cD_{{A}},\cD_{{B}}\}&=&
\cT_{ {A}{B} }{}^{{C}}\cD_{{C}}
+\hf \cR_{{A} {B}}{}^{{cd}}M_{{cd}}~,
\label{algebra-0}
\eea
where $T_{ {A}{B} }{}^{{C}}$ and $R_{{A} {B}}{}^{{cd}}$ are
the torsion and curvature tensors, respectively. 
To describe supergravity, the covariant derivatives 
have to obey certain torsion constraints \cite{GGRS} such that 
the algebra \eqref{algebra-0} takes the form \cite{KLT-M11} 
\bsubeq
\bea
\{\cD_\a,\cD_\b\}&=&
2\ri\cD_{\a\b}
-4\ri \cS M_{\a\b}
~,~~~~~~~~~
\label{N=1alg-1}
\\
{[}\cD_{a},\cD_\b{]}
&=& (\g_a)_\b{}^{\g}\Big[
\cS\cD_{\g}
+\ri 
\cC_{\g\d\r}M^{\d\r} \Big]
-\frac{2}{3}\Big[
\cD_{\b}\cS\d_a^c
-2\ve_{ab}{}^{c}(\g^b)_{\b\g}\cD^{\g}\cS\Big]M_c
~,~~~~~~~~~
\label{N=1alg-3/2-2}
\\
{[}\cD_{a},\cD_b{]}
&=&
\ve_{abc}\Big{\{}
 \Big{[}\frac{1}{2}(\g^c)_{\a\b}\cC^{\a\b\g}
-\frac{2\ri}{3}(\g^c)^{\b\g}\cD_{\b}\cS\Big{]}\cD_\g
\non\\
&&
~~~~~~
+\Big{[}
\frac{1}{2}(\g^c)^{\a\b}(\g^d)^{\g\d}\cD_{(\a}\cC_{\b\g\d)}
+\Big(
\frac{2\ri}{3}\cD^2\cS
+4\cS^2\Big)\eta^{cd}
\Big{]}M_d
\Big{\}}
~.
~~~~~~~~~~~~
\label{N=1alg-2}
\eea
\esubeq
Here the scalar $\cS$ and the symmetric spinor $\cC_{\a\b\g}=\cC_{(\a\b\g)}$ 
are real.
The dimension-2 Bianchi identities imply that 
\bea
\cD_{\a}\cC_{\b\g\d}&=&
\cD_{(\a}\cC_{\b\g\d)}
+ \ve_{\a(\b}\cD_{\g\d)}\cS \quad \Longrightarrow \quad 
\cD^\g \cC_{\a\b\g} = \frac{4 }{3} \cD_{\a\b} \cS
~.
\eea
We  use the notation $\cD^2 := \cD^\a \cD_\a$.

The algebra of covariant derivatives is invariant under 
the following super-Weyl transformations \cite{ZP88,ZP89,LR-brane}
\bsubeq \label{2.10}
\bea
\d_\s\cD_\a&=&
\hf \s\cD_\a + \cD^{\b}\s M_{\a\b}
~,
\\
\d_\s\cD_a&=&
\s\cD_a
+\frac{\ri}{ 2}(\g_a)^{\g\d}\cD_{\g} \s\cD_{\d}
+\ve_{abc}\cD^b\s M^{c}
~,
\eea
\esubeq
with the parameter $\s$ being a real unconstrained superfield, provided 
the torsion superfields transform as
\bea
\d_\s\cS&=&\s\cS-\frac{\ri}{4}  \cD^2\s~,~~~~~~
\d_\s \cC_{\a\b\g}=\frac{3}{2}\s \cC_{\a\b\g}-\frac{\ri}{2}  \cD_{(\a\b}\cD_{\g)}\s
~.
\label{sW}
\eea

 The $\cN=1$ supersymmetric extension of the 
 Cotton tensor \eqref{Cotton33} was constructed in  \cite{KT-M12}. 
 It is given by the expression
 \bea
\cW_{\a\b\g} = \left(\frac{\ri }{2} \cD^2 +4\cS\right) \cC_{\a\b \g} 
+  \ri \cD_{(\a\b} \cD_{\g)} \cS 
\label{super-Cotton}
~.
\eea
The super-Weyl transformation of $\cW_{\a\b\g}$ proves to be
\bea
\d_\s \cW_{\a\b\g} = \frac{5}{2} \s \cW_{\a\b\g}~.
\eea
It can be shown \cite{BKNT-M1} that the curved superspace is conformally flat if and 
only if $\cW_{\a\b\g}=0$.


\subsection{Conformal gauge superfields}

A real tensor superfield ${\mathfrak H}_{\a(n) } $ is said to be a conformal 
gauge supermultiplet
if (i)  it is super-Weyl primary of dimension $(1-{n}/{2})$, 
\bea
\d_\s {\mathfrak H}_{\a(n)} = \big( 1-\frac{n}{2}\big) \s {\mathfrak H}_{\a(n)}~;
\label{5.11}
\eea
and (ii) it is defined modulo gauge transformations of the form
\bea
\d_\l {\mathfrak H}_{\a(n) } =\ri^n \cD_{(\a_1} \l_{\a_2 \dots \a_n) }~,
\label{5.122}
\eea
with the  gauge parameter $\l_{\a(n-1)}$
being real but otherwise unconstrained.
The super-Weyl weight of ${\mathfrak H}_{\a(n) } $, given by  $(1-{n}/{2})$,  is uniquely 
fixed by requiring $\l_{\a(n-1)}$ and $\d_\l {\mathfrak H}_{\a(n) }$ to be 
super-Weyl primary.

Starting  with ${\mathfrak H}_{\a(n) } $ one can construct its descendant,
${\mathfrak W}_{\a(n)} ( {\mathfrak H})$, 
defined uniquely, modulo a normalisation, 
by the following the properties: 
\begin{enumerate}

\item
${\mathfrak W}_{\a(n)}$ is of the form $\cA {\mathfrak H}_{\a(n)}$, 
where $\cA$ is a linear differential operator involving $\cD_A$,
the torsion tensors
  $\cC_{\a\b\g}$ and $\cS$
 and their covariant derivatives.

\item
${\mathfrak W}_{\a(n)}$ is super-Weyl primary of weight $(1+n/2)$, 
\bea
\d_\s {\mathfrak W}_{\a(n)} = \big(1+\frac{n}{2} \big)  \s {\mathfrak W}_{\a(n)}~.
\label{5.14}
\eea

\item
The gauge variation of ${\mathfrak W}_{\a(n)}$ vanishes 
if the superspace is conformally flat,
\bea
\d_\l {\mathfrak W}_{\a(n)} = O\big( \cW_{(3)}\big)~,
\label{5.15}
\eea 
where $\cW_{(3)}$ is the super-Cotton tensor \eqref{super-Cotton}.

\item
 ${\mathfrak W}_{\a(n)}$ is divergenceless if the superspace is conformally flat,
\bea
\cD^{\b} {\mathfrak W}_{\b \a(n-1)} = O\big( \cW_{(3)}\big)~.
\label{5.16}
\eea 
Here $O\big( \cW_{(3)}\big)$ stands for contributions containing the super-Cotton tensor and
its covariant derivatives.
\end{enumerate}

As a simple example, we consider a U(1) vector multiplet coupled to supergravity,
which corresponds to the $n=1$ case. 
This multiplet is described by a real spinor prepotential 
${\mathfrak H}_\a $ which is super-Weyl primary of weight $1/2$ and is defined 
modulo gauge transformations $\d_\l  {\mathfrak H}_\a = \ri \cD_\a \l$, 
where the gauge parameter $\l$ is an unconstrained real superfield. 
The required super-Weyl primary descendant of weight $3/2$ 
is given by 
\bea
{\mathfrak W}_\a =-\frac{\ri}{2} \cD^\b \cD_\a {\mathfrak H}_\b
-2\cS {\mathfrak H}_\a 
\eea
and proves to be gauge invariant, 
\bea
\d_\z  {\mathfrak W}_\a =0~.
\eea
The field strength obeys the Bianchi identity 
\bea
\cD^\a {\mathfrak W}_\a =0~.
\eea
For $n>1$ the right-hand sides of \eqref{5.15} and \eqref{5.16} are non-vanishing.

Suppose that our background  curved superspace $\cM^{3|2}$
is conformally flat, 
\bea
\cW_{\a(3)}=0~.
\eea
Then the tensor superfield ${\mathfrak W}_{\a(n)}$ 
is gauge invariant and conserved, 
\begin{subequations}
\bea
\d_\l {\mathfrak W}_{\a(n)}&=&0~,\\ 
\cD^{\b} {\mathfrak W}_{\b\a(n-1)} &=&0~. 
\eea
\end{subequations}
These properties and the super-Weyl transformation laws
\eqref{5.11} and  \eqref{5.14} imply
that the action\footnote{The super-Weyl transformation of the superspace 
integration measure is $\d_\s E = -2\s E$.} 
\bea
{\mathbb S}_{\rm{SCS}}^{(n)} [ {\mathfrak H}_{(n)}] 
= - \frac{\ri^n}{2^{\left \lfloor{n/2}\right \rfloor +1}}
   \int \rd^{3|2}z \, E\,
 {\mathfrak H}^{\a(n)} 
{\mathfrak W}_{\a(n) }( {\mathfrak H}) ~, \qquad E^{-1} = {\rm Ber} (E_A{}^M)
\label{5.21}
\eea
is gauge and super-Weyl invariant, 
\bea
\d_\l {\mathbb S}_{\rm{SCS}}^{(n)} [{\mathfrak H}_{(n)}] =0~, \qquad 
\d_\s {\mathbb S}_{\rm{SCS}}^{(n)} [ {\mathfrak H}_{(n)} ] =0~.
\eea
We now turn to constructing the linearised higher-spin 
super-Cotton tensors ${\mathfrak W}_{\a(n)}$
on such a conformally flat superspace. 


\subsection{Higher-spin super-Cotton tensor in Minkowski superspace}

In Minkowski superspace, ${\mathbb M}^{3|2}$, the higher-spin 
super-Cotton tensor \cite{K16,KT} is 
\bea
W_{\a_1 \dots \a_n} = 
\Big( -\frac{\ri}{2}\Big)^n
D^{\b_1} D_{\a_1} \dots D^{\b_n} D_{\a_n} H_{\b_1 \dots \b_n}
= W_{(\a_1 \dots \a_n )} 
~,
\label{5.17}
\eea
with $D_A = (\pa_a, D_\a)$ being the flat-superspace covariant derivatives.  
This tensor is invariant under the gauge transformation 
\bea
\d H_{\a_1\a_2  \dots \a_n } &=& \ri^n D_{(\a_1 } \z_{\a_2 \dots \a_n)} ~, 
\label{5.18}
\eea
and obeys the conservation identity 
\bea
{D}^{\b} {W}_{\b\a_1 \dots \a_{n-1}}=0 ~.
\label{5.25}
\eea
The fact that $W_{\a_1 \dots \a_n}$ 
defined by \eqref{5.17} is completely symmetric, 
is a corollary of the identities
\bea
D^\a D_\b D_\a =0 \quad \Longrightarrow \quad 
[D_\a D_\b, D_\g D_\d ]=0~. 
\eea

The normalisation in \eqref{5.17} is explained as follows.
The gauge freedom \eqref{5.18} allows us to impose a gauge condition
\bea
D^\b H_{\b \a(n-1)} =0~,
 \eea
 under which
 the expression for the super-Cotton tensor 
 simplifies,
\begin{subequations} 
\bea
 D^\b H_{\b \a_1 \dots \a_{n-1}}=0 \quad \Longrightarrow \quad 
 W_{\a(n)} = \pa_{\a_1}{}^{\b_1} \dots \pa_{\a_n}{}^{\b_n} H_{\b_1 \dots \b_n}~.
\eea
This result can be fine-tuned to
\bea
W_{\a(2s)} &=&\Box^s H_{\a(2s)}~, \\
W_{\a(2s+1)} &=&\Box^s \pa^\b{}_{(\a_1} H_{\a_2 \dots \a_{2s+1})\b }
=\Box^s \pa^\b{}_{\a_1 } H_{\a_2 \dots \a_{2s+1} \b }~,
\eea
\end{subequations}
where $s>0$ is an integer.

For completeness, we also give another representation 
for the higher-spin super-Cotton tensor  derived in \cite{K16,KT}:
\bea
&&W_{\a_1 \dots \a_n} 
:= \frac{1}{2^{n}} 
\sum\limits_{j=0}^{\left \lfloor{n/2}\right \rfloor}
\bigg\{
\binom{n}{2j}  \Box^{j}\pa_{(\a_{1}}{}^{\b_{1}}
\dots
\pa_{\a_{n-2j}}{}^{\b_{n-2j}}H_{\a_{n-2j+1}\dots\a_{n})\b_1 \dots\b_{n-2j}}~~~~
\nonumber \\
&&\qquad \qquad -\frac{\ri}{2} 
\binom{n}{2j+1}D^{2}\Box^{j}\pa_{(\a_{1}}{}^{\b_{1}}
\dots\pa_{\a_{n-2j -1}}{}^{\b_{n-2j -1}}H_{\a_{n-2j}\dots\a_{n})
\b_1 \dots \b_{n-2j -1} }\bigg\}~.~~~~~
\label{5.30}
\eea

The following higher-spin action \cite{K16,KT}
\bea
{\mathbb S}^{(n)}_{\rm SCS}[H_{(n)}] = 
- \frac{\ri^n}{2^{\left \lfloor{n/2}\right \rfloor +1}}
 \int \rd^{3|2}z \,H^{\a(n)} W_{\a(n)}(H)
\label{5.31}
\eea
is $\cN=1$ superconformal. It is  clearly invariant under the gauge transformations 
\eqref{5.18}.


\subsection{Higher-spin super-Cotton tensor in conformally flat superspaces}

Consider a curved conformally flat superspace $\cM^{3|2}$.
Locally, its covariant derivatives $\cD_A$ are related to the flat-space ones by 
\bea
\cD_\a &=&
\re^{\hf \s} \Big(D_\a + \cD^{\b}\s M_{\a\b} \Big)
~,
\\
\cD_a
&=&
\re^\s \Big( 
\pa_a + \frac{\ri}{2} (\g_a)^{\a\b}\cD_\a \s D_\b
+ 
 \pa^b \s M_{ba}
- \frac{\ri}{8} (\g_a)^{\a\b} (\cD^\g \s) D_\g \s M_{\a\b} \Big)~,
\eea
for some scale factor $\s$.
In accordance with \eqref{5.14},
the  higher-spin 
super-Cotton tensor ${\mathfrak W}_{\a(n)}$ in $\cM^{3|2}$
is related to the flat-space one, eq. \eqref{5.17} or equivalently \eqref{5.30},
by the rule
\bea
{\mathfrak W}_{\a(n)} = \re^{(1+\frac{n}{2})\s} W_{\a(n)}~. 
\eea

In general, it is a difficult technical problem to express ${\mathfrak W}_{\a(n)}$ 
in terms of the covariant derivatives $\cD_A$ and the gauge prepotential  
${\mathfrak H}_{\a(n)} =  \re^{(1-n/2)\s} H_{\a(n)}$.
As an example, 
we only give expressions for the supersymmetric photino ${\mathfrak W}_{\a}$ 
and Cottino ${\mathfrak W}_{\a(2)}$ tensors in
AdS superspace. The geometry of AdS${}^{3|2}$
is encoded in the following
algebra of covariant derivatives: 
\begin{subequations}
\bea
\big\{ {\cal D}_\a , {\cal D}_\b \big\} &=&2 \ri {\cal D}_{\a \b}  - 4\ri\cS M_{\a \b} ~,\\ 
\big[ {\cal D}_{\a \b}, {\cal D}_\g \big]&=&-2\cS \ve_{\g (\a}{\cal D}_{\b)} ~, \\ 
\big[ {\cal D}_{\a \b}, {\cal D}_{\g \delta} \big]&=&4\cS^2 (\ve_{\g (\a}{M}_{\b) \delta} 
+ \ve_{\delta (\a}{M}_{\b) \g})~,
\eea
 \end{subequations}
 with the  real parameter $\cS$ being the same as in  \eqref{3.45a}. 
 The tensors 
 ${\mathfrak W}_{\a}$ 
and ${\mathfrak W}_{\a(2)}$ 
are expressed in terms of the operator
\bea
\D^\b{}_\a := -\frac{\ri}{2} \cD^\b \cD_\a -2 \cS \d^\b{}_\a~,
\eea
with the properties 
\bea
\big[ \D^{\b_1}{}_{\a_1} , \D^{\b_2}{}_{\a_2} \big] &=&
\ve_{\a_1 \a_2} \cS (\cD^{\b_1\b_2} -2\cS M^{\b_1 \b_2} )
-\ve^{\b_1\b_2} \cS ( \cD_{\a_1\a_2}  - 2\cS M_{\a_1 \a_2} )~.
\eea
These properties follow from the identity
\bea
\cD^\b \cD_\a \cD_\b = 4\ri \cS \cD_\a \quad \implies \quad
\cD^\a \D^\b{}_\a =0~.
\label{5.37}
\eea
The expressions for ${\mathfrak W}_{\a}$ 
and  ${\mathfrak W}_{\a(2)}$ are:
\begin{subequations} 
\bea
{\mathfrak W}_{\a} &:=& \D^\b{}_\a {\mathfrak H}_{\b}~,\\
{\mathfrak W}_{\a_1 \a_2} &=& \D^{\b_1}{}_{(\a_1}  \D^{\b_2}{}_{\a_2)} 
{\mathfrak H}_{\b_1 \b_2} -2 \cS  \D^{\b}{}_{(\a_1}  {\mathfrak H}_{\a_2) \b}~.
\eea
\end{subequations} 


\subsection{Massive supersymmetric higher-spin theories in AdS superspace}

Massive supersymmetric higher-spin actions in AdS involve different 
massless sectors depending on the value of superspin. 
\begin{subequations} \label{5.39}
\bea
{\mathbb S}_{\rm massive}^{(2s)}
&=&\lambda {\mathbb S}_{\rm{SCS}}^{(2s)}[{\mathfrak H}_{(2s)}]
+\mu^{2s-1}{\mathbb S}_{\rm{FO}}^{(2s)} [{\mathfrak H}_{(2s)},{\mathfrak Y}_{(2s-2)}]~,
\label{5.39a}
\\
{\mathbb S}_{\rm massive}^{(2s+1)}
&=&\lambda {\mathbb S}_{\rm{SCS}}^{(2s+1)}[{\mathfrak H}_{(2s+1)}]
+\mu^{2s-1}{\mathbb S}_{\rm{SO}}^{(2s+1)}
[{\mathfrak H}_{(2s+1)},{\mathfrak X}_{(2s-2)}]
\label{5.39b}
\eea
\end{subequations}

\subsubsection{First-order massless actions}

We introduce a gauge theory described  by a reducible gauge superfield 
${\bm \cH}_{\b , \a_1 \dots \a_{n-1}} ={\bm \cH}_{\b , (\a_1 \dots \a_{n-1})}$.
This superfield is defined modulo
gauge transformations of the form
\bea
\d {\bm \cH}_{\b , \a_1 \dots \a_{n-1}} = \ri^n \cD_\b \l_{\a_1 \dots \a_{n-1}}~.
\label{5.40}
\eea
A supersymmetric gauge-invariant action of lowest order in derivatives is 
\bea
{\mathbb S}^{(n)}_{\rm FO} = \frac{\ri^{n+1}}{2^{\lceil (n+1)/2\rceil}}
\int \rd^{3|2} z \, E\, 
{\bm \cH}^{\b , \a_1 \dots \a_{n-1}} \Big(
 \cD^\g \cD_\b 
  -4\ri \cS \d^\g{}_\b \Big)
{\bm  \cH}_{\g , \a_1 \dots \a_{n-1}}
   ~.
\label{5.41}
\eea
The gauge invariance of ${\mathbb S}^{(n)}_{\rm FO}$ 
follows from the identity \eqref{5.37}.
Our action \eqref{5.41} is a  higher-spin AdS extension of the model for the massless
gravitino multiplet ($n=2$) in Minkowski superspace proposed by Siegel \cite{Siegel}
(see also \cite{GGRS}).

The gauge superfield ${\bm \cH}_{\b , \a_1 \dots \a_{n-1}}$ can be decomposed 
 into irreducible ${\rm SL} (2,{\mathbb R})$ superfields
\bea
{\bm \cH}_{\b , \a_1 \dots \a_{n-1}} = {\mathfrak H}_{\b \a_1 \dots \a_{n-1} }
+
 \sum_{k=1}^{n-1} \ve_{\b \a_k} {\mathfrak Y}_{\a_1 \dots {\hat \a}_k \dots \a_{n-1}}~,
\eea
where ${\mathfrak H}_{\a (n) } $ and ${\mathfrak Y}_{\a (n-2)}$ 
are completely symmetric 
tensor superfields. Then the gauge transformation \eqref{5.40} turns into
\begin{subequations}
\bea
\d {\mathfrak H}_{\a (n)} &=& \ri^n \cD_{(\a_1 } \l_{\a_2 \dots \a_n)} ~, \\
\d {\mathfrak Y}_{\a (n-2) } &=& \frac{\ri^n}{n} \cD^\b \l_{\b \a_1 \dots \a_{n-2}}~.
\eea
\end{subequations}
The supersymmetric gauge-invariant action takes the form 
\bea
{\mathbb S}^{(n)}_{\rm FO} &=& 
\frac{\ri^{n+1}}{2^{\lceil (n+1)/2\rceil}}
\int \rd^{3|2} z \, E\,\Big\{  
{\mathfrak H}^{\b  \a (n-1)} \cD^\g \cD_\b {\mathfrak H}_{\g   \a (n-1)}  
 +2\ri (n-1) {\mathfrak Y}^{\a (n-2)} \cD^{\b\g} {\mathfrak H}_{ \b \g \a (n-2) } \non \\
&&
+ (n-1) \Big( {\mathfrak Y}^{\a (n-2) }\cD^2 {\mathfrak Y}_{\a (n-2)} 
+(-1)^n 
(n-2)\cD_\b {\mathfrak Y}^{\b \a (n-3) }
\cD^\g {\mathfrak Y}_{\g \a (n-3)} \Big) \non \\
&& -4\cS \ri \Big( {\mathfrak H}^{\a(n) } {\mathfrak H}_{\a(n) }
+n(n-1) {\mathfrak Y}^{\a(n-2)} {\mathfrak Y}_{\a(n-2)} \Big)
\Big\} ~.
\label{5.44}
\eea
When $n$ is even, $n=2s$, this action is the unique gauge-invariant AdS extension 
of the massless integer superspin action of \cite{KT}.


\subsubsection{Second-order massless actions}

The massless half-integer superspin action in AdS is 
\bea
{\mathbb S}_{\rm{SO}}^{(2s+1)}
&=& \Big(- \hf \Big)^s   \int \rd^{3|2} z \, E\,\bigg\{ 
-\frac{\ri}{2} {\mathfrak H}^{\a(2s+1)} {\mathbb Q} {\mathfrak H}_{\a(2s+1)}
-\frac{\ri}{8} \cD_\b {\mathfrak H}^{\b \a(2s)} \cD^2 \cD^\g  {\mathfrak H}_{\g \a(2s)} \non \\
&&+ \frac{\ri}{4} s \cD_{\b\g} {\mathfrak H}^{\b\g \a(2s-1)} 
\cD^{\r\l} {\mathfrak H}_{\r\l \a(2s-1)}
-\hf (2s-1) {\mathfrak X}^{\a(2s-2)} \cD^{\b\g} \cD^\d {\mathfrak H}_{\b\g \d \a(2s-2)} \non \\
&&+\frac{\ri}{2}  (2s-1)\Big[ {\mathfrak X}^{\a(2s-2)} \cD^2 {\mathfrak X}_{\a(2s-2)}
- \frac{s-1}{s} \cD_\b {\mathfrak X}^{\b\a(2s-3)} \cD^\g {\mathfrak X}_{\g \a(2s-3)}\Big]
\non \\
&& +\ri s \cS {\mathfrak H}^{\b \a(2s)}  \cD_\b{}^\g  {\mathfrak H}_{\g \a(2s)}
+\hf (s+1) {\cS} {\mathfrak H}^{\a(2s+1)} \cD^2 {\mathfrak H}_{\a(2s+1)}
 \label{5.45} \\
&&+\ri s (2s-3) {\cS}^2 {\mathfrak H}^{\a(2s+1)} {\mathfrak H}_{\a(2s+1)}
+ \frac{(2s-1)(s^2 -3s -2)}{s} {\cS} X^{\a(2s-2)} X_{\a(2s-2)}\bigg\}
~,~~~
\non
\eea
where $\mathbb Q$ is the quadratic Casimir operator \eqref{2.155}.
One can express $\mathbb Q$ in the form 
\bea
{\mathbb Q } = \cD^a \cD_a -\ri {\cS} \cD^2 
+2 {\cS} \cD^{\a \b} M_{\a\b} -2 {\cS}^2 M^{\a\b} M_{\a\b}~.
\eea
The action \eqref{5.45} is invariant under the gauge transformations
\begin{subequations}
\bea
\d {\mathfrak H}_{\a(2s+1)} &=& \ri \cD_{(\a_1} \l_{\a_2 \dots \a_{2s+1} )} ~,\\
\d {\mathfrak X}_{\a(2s-2)} &=& \frac{s}{2s+1} \cD^{\b\g} \l_{\b\g \a_1 \dots \a_{2s-2} }~.
\eea
\end{subequations}  
The action \eqref{5.45} is the unique gauge-invariant AdS extension 
of the massless half-integer superspin action of \cite{KT}.


\subsection{From AdS superspace to AdS space}

To conclude this section, we briefly discuss the key aspects of component 
reduction for supersymmetric field theories formulated in AdS superspace,
AdS$^{3|2}$.
In general, the action functional of such a theory 
is given by 
\bea
S=  \int \rd^{3|2} z \, E\,\cL ~,
\label{5.47}
\eea
where the Lagrangian $\cL$ is a scalar superfield. 
In accordance with the general formalism described in section 6.4 of \cite{BK}, 
the  isometry transformations of AdS$^{3|2}$ are generated by the Killing vector fields
$\x^A E_A$ which are defined to obey the master equation \cite{KLT-M12}
\bea
\big[\x+\hf \L^{bc}M_{bc},\cD_{A} \big]=0~,\qquad
\x:=\x^B \cD_B =
\x^{b} \cD_{b}+\x^\b \cD_\b~, 
\label{N=1-killings-0}
\eea
for some Lorentz superfield parameter   $\L^{bc}= -\L^{cb}$. 
An infinitesimal isometry transformation acts on 
a tensor superfield $T$ as 
\bea
\d_\x T = \big(\x+\hf \L^{bc}M_{bc} \big) T ~.
\eea
The action \eqref{5.47} is 
invariant under the isometry group of  AdS$^{3|2}$.

As shown in \cite{KLT-M12}, the parameters in \eqref{N=1-killings-0}
obey the following Killing equations:
\begin{subequations}\label{5.50}
\bea
\cD_\a \x_\b &=&\hf \L_{\a\b} + \cS \x_{\a\b} = \cD_\b \x_\a~, \\
\cD_\b \x^{\b \a} +6\ri \x^\a &=&0~, \qquad 
\cD_\b \L^{\b \a} +12\cS \ri \x^\a =0~, \\
\cD_{(\a} \x_{\b\g)} &=&0~, \qquad \cD_{(\a} \L_{\b\g)}=0~,
\eea
\end{subequations}
which imply 
\begin{subequations}\label{5.51}
\bea
 \cD_a \x_b + \cD_b \x_a &=& 0~, \label{5.51a}\\
\cD^2  \x_\a -12 \ri \cS \x_\a &=&0~, \\
\cD_{\a\b} \x^\b  +2\cS \x_\a &=&0~. \label{5.51c}
\eea
\end{subequations}
Equation \eqref{5.51a} tells us that $\x_a$ is a Killing vector, while 
\eqref{5.51c} means that $\x_\a$ is a Killing spinor. 
The component form of the action \eqref{5.47} is computed using the formula
(see also \cite{KT-M12})
\bea
S =  \frac{1}{4} \int \rd^3 x \, e\,\big(\ri \cD^2 +8 \cS \big) \cL \big|~.
\label{5.52}
\eea
Here and 
in what follows, the $\q$-independent component $T |_{\q=0}$
of a superfield $T(x,\q)$ will simply be denoted $T|$. To complete the formalism 
of component reduction, we only need the following relation 
\bea
\big(\cD_a T \big) \big| = \nabla_a T|~,
\eea
where $\nabla_a$ is the standard torsion-free covariant derivative of AdS space.
Making use of the AdS transformation law
$ \d_\x \cL = \x \cL$ in conjunction with the identities \eqref{5.50} and 
\eqref{5.51}, one may check that the action \eqref{5.52} is invariant 
under arbitrary isometry transformations of the AdS superspace.


\section{Supersymmetric higher-spin actions in components}

In this section we will describe the component structure of the supersymmetric 
higher-spin theories introduced in the previous section.
Our analysis will be restricted to the flat-superspace case. 
As in \cite{KT}, the integration measure\footnote{This 
definition implies that $ \int \rd^{3|2} z \, V = \int \rd^3 x\, F$, for any scalar 
superfield $V(x,\q)  =\dots + \ri \q^2 F(x)$.} for $\cN=1$ Minkowski superspace
is defined as follows:
\bea
\int \rd^{3|2} z \, L= \frac{\ri}{4} \int \rd^3 x \, D^2 L\big|_{\q=0}~.
\label{6.1}
\eea

\subsection{Superconformal higher-spin action}

We start by reducing the superconformal higher-spin action 
\eqref{5.31} to components. 
The gauge freedom \eqref{5.18} can be used to impose a Wess-Zumino gauge
\bea
H_{\a_1 \dots \a_n} |=0~,\qquad
D^\b H_{\b \a_1 \dots \a_{n-1}} | =0~.
\label{6.2}
\eea
In this gauge, there remain two independent component fields
\bea
h_{\a_1 \dots \a_{n+1}}:= \ri^{n+1} D_{(\a_1} H_{\a_2 \dots \a_{n+1} )} \big|~, \qquad
h_{\a_1 \dots \a_n}  := -\frac{\ri}{4} D^2 H_{\a_1 \dots \a_n}\big|~.
\eea
Due to the conservation equation \eqref{5.25}, the higher-spin super-Cotton tensor 
\eqref{5.30} also has two independent components, which we define as
\bea
C_{\a_1 \dots \a_n}  := W_{\a_1 \dots \a_n}\big|~,\qquad
C_{\a_1 \dots \a_{n+1}}:= \ri^{n+1} D_{(\a_1} W_{\a_2 \dots \a_{n+1} )} \big|~.
\eea
The field strengths $C_{\a(n)}$ and $C_{\a(n+1)}$ are given in terms of 
the gauge potentials $h_{\a(n)}$ and $h_{\a(n+1)}$, respectively, 
according to  eq. \eqref{2.31}. To prove this statement for 
$C_{\a(n+1)}$, one has to use the identity 
\bea
\binom{n}{2j} +\binom{n}{2j+1} =\binom{n+1}{2j+1} ~.
\eea
Reducing the action \eqref{5.31} to components gives 
\bea
{\mathbb S}^{(n)}_{\rm SCS}[H_{(n)}] =
S^{(n)}_{\rm CS}[h_{(n)}]  + S^{(n+1)}_{\rm CS}[h_{(n+1)}] ~, 
\label{6.6}
\eea
where the conformal higher-spin action $S^{(n)}_{\rm CS}[h_{(n)}]$ 
is defined by eq. \eqref{3.39}.

In the gauge \eqref{6.2}, the residual gauge freedom is characterised by the conditions
\bea
D_{(\a_1 } \l_{\a_2 \dots \a_n )}|=0~, \qquad 
D^2 \l_{\a_1 \dots \a_{n-1}} |= -2\ri \frac{n-1}{n+1} \pa^\b{}_{(\a_1} 
\l_{\a_2 \dots \a_{n-1} )\b} |~.
\eea
At the component level, the remaining independent gauge transformations 
are generated by $\z_{\a(n-1)} \propto \l_{\a(n-1)}\big|$ and 
$\z_{\a(n-2)} \propto \ri^n D^\b \l_{\b\a(n-2)}\big|$.


\subsection{Massless first-order model}

We now turn to 
working out the component structure of the first-order model  \eqref{5.44}
in the flat-superspace limit. In Minkowski superspace, the action can be written in 
the form
\bea
{\mathbb S}^{(n)}_{\rm FO} &=& 
\frac{\ri^{n+1}}{2^{\lceil (n+1)/2\rceil}}
\int \rd^{3|2} z \, \Big\{ \ri H^{\b  \a_1 \dots \a_{n-1}} \pa_\b{}^\g 
H_{\g  \a_1 \dots \a_{n-1}}  
+\hf H^{\a_1 \dots \a_n}D^2 H_{\a_1 \dots \a_n}  \non \\
&& +2\ri (n-1) Y^{\a_1 \dots \a_{n-2}} \pa^{\b\g} H_{ \b \g \a_1 \dots \a_{n-2} } 
+ (n-1) Y^{\a_1 \dots \a_{n-2} }D^2 Y_{\a_1 \dots \a_{n-2}} \non \\
&& +(-1)^n (n-1)(n-2)D_\b Y^{\b \a_1 \dots \a_{n-3} }
D^\g Y_{\g \a_1 \dots \a_{n-3}}\Big\} ~.
\label{6.88}
\eea
It is invariant under the gauge transformations
\begin{subequations}
\bea
\d H_{\a_1\a_2  \dots \a_n } &=& \ri^n D_{(\a_1 } \l_{\a_2 \dots \a_n)} ~, \label{2.3a}\\
\d Y_{\a_1  \dots \a_{n-2} } &=& \frac{\ri^n}{n} D^\b \l_{\b \a_1 \dots \a_{n-2}}~,
\eea
\end{subequations}
with the gauge parameter $\l_{\a(n-1)}$ being a real unconstrained superfield. 
When $n$ is even, $n=2s$, the action \eqref{6.88} describes the massless
integer superspin model of \cite{KT}.


The gauge freedom allows us to choose a Wess-Zumino gauge
\bea
H_{\a_1 \dots \a_n} |=0~,\qquad
D^\b H_{\b \a_1 \dots \a_{n-1}} | =0~, \qquad
Y_{\a_1 \dots \a_{n-2}} |= 0~.
\eea
Then, the residual gauge freedom is characterised by the conditions
\bea
D_{\a_1 } \l_{\a_2 \dots \a_n }|=0~, \qquad 
D^2 \l_{\a_1 \dots \a_{n-1}} |= -2\ri \frac{n-1}{n+1} \pa^\b{}_{(\a_1} 
\l_{\a_2 \dots \a_{n-1} )\b} |~.
\eea
These conditions imply that there remains only one independent gauge parameter 
at the component level. We define it as 
\bea
\z_{\a_1 \dots \a_{n-1}} (x) := (-1)^{n+1} \l_{\a_1 \dots \a_{n-1}}|~.
\eea
We define the component fields as 
\begin{subequations}
\bea
h_{\a_1 \dots \a_{n+1}}&:= &\ri^{n+1} D_{(\a_1} H_{\a_2 \dots \a_{n+1} )} |~, \\
h_{\a_1 \dots \a_n} & :=& -\frac{\ri}{4} D^2 H_{\a_1 \dots \a_n}|~, \\
y_{\a_1 \dots \a_{n-1}}&:= & \frac{\ri^{n+1}}{2n} D_{(\a_1} Y_{\a_2 \dots \a_{n-1} )} |~,
\qquad y_{\a_1 \dots \a_{n-3}}:= \ri^{n+1} D^\b Y_{\b\a_1 \dots \a_{n-3} } |~,\\
Z_{\a_1 \dots \a_{n-2}} & :=& \frac{\ri}{4} D^2 Y_{\a_1 \dots \a_{n-2}}|~.
\eea
\end{subequations}
Their gauge transformation laws are
\begin{subequations}
\bea
\d h_{\a_1 \dots \a_{n+1} }&=& \pa_{(\a_1 \a_2} \z_{\a_3 \dots \a_{n+1})} ~,\\
\d y_{\a_1 \dots \a_{n-1} } &=&\frac{1}{n+1}  \pa^\b{}_{(\a_1 } \z_{\a_2 \dots \a_{n-1} )\b} ~,\\
\d y_{\a_1 \dots \a_{n-3} }&=& \pa^{\b \g} \z_{\b\g \a_1 \dots \a_{n-3}} ~, \\
\d h_{\a_1 \dots \a_n}&=&0~, \\
 \d Z_{\a_1 \dots \a_{n-2} } &=&0~.
\eea
\end{subequations}
Direct calculations of the component action give
\bea
{\mathbb S}^{(n)}_{\rm FO}
&=&  
\frac{\text{i}^n}{2^{\lceil n/2\rceil}}
\int \rd^3 x \, \Big\{ h^{\a_1 \dots \a_n} h_{\a_1 \dots \a_n}
+Z^{\a_1 \dots \a_{n-2}} Z_{\a_1 \dots \a_{n-2}} \Big\} \non \\
&+&
\frac{\ri^{n+1}}{2^{\lceil (n+1)/2\rceil}}
\int \rd^3 x \, \Big\{ h^{\b \a_1 \dots \a_n} \pa_\b{}^\g h_{\g \a_1 \dots \a_n}
+2(n-1) y^{\a_1 \dots \a_{n-1}} \pa^{\b\g} h_{\b \g \a_1 \dots \a_{n-1}}  \non \\
&&  + 4(n-1) y^{\b \a_1 \dots \a_{n-2}} \pa_\b{}^\g y_{\g \a_1 \dots \a_{n-2}}
+ \frac{2(n-2)(n+1)}{n}y^{\a_1 \dots \a_{n-3}} \pa^{\b\g} y_{\b \g \a_1 \dots \a_{n-3}} \non \\
&&- \frac{(n-2)(n-3)}{n(n-1)} y^{\b \a_1 \dots \a_{n-4}} \pa_\b{}^\g h_{ \g \a_1 \dots \a_{n-4}}
\Big\}~.
\eea
The fields $h_{\a(n)}$ and $Z_{\a(n-2)}$ appear in the action without derivatives.
This action can  be rewritten in the form 
\bea
{\mathbb S}^{(n)}_{\rm FO}
&=&  
\frac{\text{i}^n}{2^{\lceil n/2\rceil}}
\int \rd^3 x \, \Big\{ h^{\a (n)} h_{\a (n)}
+Z^{\a (n-2)} Z_{\a (n-2)} \Big\} 
+ S_{\rm{FF}}^{(n+1)}
[{h}_{(n+1)},{y}_{(n-1)},
{ y}_{(n-3)}]~,~~~
\label{6.16}
\eea
where $S_{\rm{FF}}^{(n+1)}$ is the flat-space version of \eqref{33.41}, 
eg. \eqref{B.5},
with $n$ replaced by $(n+1)$.


\subsection{Massive integer superspin action}

We are now prepared to read off the component form of a massive 
integer superspin action that is obtained 
from \eqref{5.39a} in the flat-superspace limit,
\bea
{\mathbb S}_{\rm massive}^{(2s)}
=\lambda {\mathbb S}_{\rm{SCS}}^{(2s)}[{H}_{(2s)}]
+\mu^{2s-1}{\mathbb S}_{\rm{FO}}^{(2s)} [{H}_{(2s)},{ Y}_{(2s-2)}]~.
\label{6.17}
\eea
Choosing $n=2s$ in the component actions \eqref{6.6} and \eqref{6.16} 
gives
\bea
{\mathbb S}_{\rm massive}^{(2s)} &=&
\l S^{(2s)}_{\rm CS}[h_{(2s)}]  
+\hf \Big( -\hf\Big)^s \mu^{2s-1}
\int \rd^3 x \,  h^{\a (2s)} h_{\a (2s)} \non \\
&&+ \l S^{(2s+1)}_{\rm CS}[h_{(2s+1)}] + 
\mu^{2s-1} S_{\rm{FF}}^{(2s+1)}
[{h}_{(2s+1)},{y}_{(2s-1)},{ y}_{(2s-3)}] \non \\
&&+\hf \Big( -\hf \Big)^s 
\mu^{2s-1} \int \rd^3 x \,  Z^{\a (2s-2)} Z_{\a (2s-2)} ~.
\label{6.18}
\eea
It is seen that the $Z_{\a(2s-2)}$ field appears only in the third line of \eqref{6.18}
and without derivatives, and thus  $Z_{\a(2s-2)}$ is an auxiliary field.
Next, the expression in the second line of \eqref{6.18} constitutes the massive 
gauge-invariant spin-$(s+\hf)$ action \eqref{3.8}. The two terms in the first 
line of \eqref{6.18} involve the $h_{\a(2s)}$ field.
Unlike $ S^{(2s)}_{\rm CS}[h_{(2s)}]$, the second mass-like term is not gauge invariant.
However, the action 
\bea
\widetilde{S}_{\rm massive}^{(2s)} 
=\l S^{(2s)}_{\rm CS}[h_{(2s)}]  
+\hf \Big( -\hf\Big)^s \mu^{2s-1}
\int \rd^3 x \,  h^{\a (2s)} h_{\a (2s)} 
\label{6.19}
\eea
does describe a massive  spin-$s$ field on-shell.
Indeed, the equation of motion is 
\bea
\l C_{\a(2s)} + \mu^{2s-1}  h_{\a (2s)}=0~.
\label{6.20}
\eea
Since $C_{\a(2s)} $ is divergenceless, eq. \eqref{3.334}, the equation of motion implies
that $h_{\a(2s)} $ is divergenceless, eq. \eqref{3.38}.
As a consequence,  $C_{\a(2s)} $ takes the simple form given by 
\eqref{3.41a}, and the above equation of motion turns into 
(compare with eq. \eqref{3.40})
\bea
\l \Box^{s-1}\partial^{\beta}{}_{\a_1}h_{\a_2\dots\a_{2s}\beta} 
+ \mu^{2s-1}  h_{\a (2s)}=0~,
\label{6.21}
\eea
which implies 
\bea
\Big(\Box^{2s-1}-(m^2)^{2s-1}\Big) h_{\alpha(2s)}=0~,
\qquad m:=\Big|\frac{\mu }{\lambda^{1/(2s-1)}}\Big|~,
\label{6.22}
\eea
and should be compared with \eqref{3.44}. 
Since the polynomial equation $z^{2s-1}-1=0$ has only one real root, 
$z=1$, we conclude that \eqref{6.22} leads to the Klein-Gordon 
equation  \eqref{3.26}. As a result, the higher-derivative equation \eqref{6.21}
reduces to the first-order one, eq.  \eqref{3.27}.

The above component analysis clearly demonstrates that the model \eqref{6.17}
describes a single massive supermultiplet subject to the equations
\eqref{214a} and \eqref{214b}  with $n=2s$ on the mass shell.
The superfield proof was provided in \cite{KT}.


\subsection{Massless second-order model}

Finally we consider the massless half-integer superspin model 
describe by the action \cite{KT}
\bea
{\mathbb S}_{\rm{SO}}^{(2s+1)}
&=& \Big(- \hf \Big)^s   \int \rd^{3|2} z \, \Big\{ 
-\frac{\ri}{2} H^{\a(2s+1)} \Box H_{\a(2s+1)}
-\frac{\ri}{8} D_\b H^{\b \a(2s)} D^2 D^\g  H_{\g \a(2s)} \non \\
&&+ \frac{\ri}{4} s \pa_{\b\g} H^{\b\g \a(2s-1)} \pa^{\r\l} H_{\r\l \a(2s-1)}
-\hf (2s-1) X^{\a(2s-2)} \pa^{\b\g} D^\d H_{\b\g \d \a(2s-2)} \non \\
&&+\frac{\ri}{2}  (2s-1)\Big[ X^{\a(2s-2)} D^2 X_{\a(2s-2)}
- \frac{s-1}{s} D_\b X^{\b\a(2s-3)} D^\g X_{\g \a(2s-3)}\Big]\Big\}~.~~
\label{7.2}
\eea
It is invariant under the following gauge transformations
\begin{subequations}
\bea
\d H_{\a(2s+1)} &=& \ri D_{(\a_1} \l_{\a_2 \dots \a_{2s+1} )} ~,\\
\d X_{\a(2s-2)} &=& \frac{s}{2s+1} \pa^{\b\g} \l_{\b\g \a_1 \dots \a_{2s-2} }~.
\eea
\end{subequations} 

The gauge freedom allows us to choose a Wess-Zumino gauge of the form 
\bea
H_{\a(2s+1)} \big|=0~, \qquad
D^\b H_{\b \a(2s)}\big|=0~.
\label{7.3}
\eea
To preserve these conditions, the residual gauge symmetry has to be constrained by 
\bea
 D_{(\a_1} \l_{\a_2 \dots \a_{2s+1} )} \big|=0~,\qquad
 D^2 \l_{\a(2s)}\big| = -\frac{2\ri s}{s+1} \pa^\b{}_{(\a_1} \l_{\a_2 \dots \a_{2s})\b}\big|~.
 \eea
Under the gauge conditions imposed, 
the independent component fields of $H_{\a(2s+1)} $ can be chosen as 
\bea
h_{\a(2s+2)} := -D_{(\a_1} H_{\a_2 \dots \a_{2s+2})} \big|~, \qquad
h_{\a(2s+1)}:= \frac{\ri}{4} D^2 H_{\a(2s+1)} \big|~.
\eea
The remaining independent component parameters of $\l_{\a(2s)}$ can be chosen as 
\bea
\z_{\a(2s)}:= \l_{\a(2s)}\big|~, \qquad
\x_{\a(2s-1)}:= -\ri \frac{s}{2s+1} D^\b \l_{\b \a(2s-1)}\big|~.
\eea
The gauge transformation laws of $h_{\a(2s+2)}$ and $h_{\a(2s+1)}$
can be shown to be 
\begin{subequations}
\bea
\d_\z h_{\a(2s+2)} &=& \pa_{(\a_1 \a_2} \z_{\a_3 \dots \a_{2s+2} )}~, \label{7.7a} \\
\d_\x h_{\a(2s+1)} &=& \pa_{(\a_1 \a_2} \x_{\a_3 \dots \a_{2s+1} )}~. \label{7.7b} 
\eea 
\end{subequations}
We now define the component fields of $X_{\a(2s-2)}$ as follows:
\begin{subequations}
\bea
y_{\a(2s-2)} &:=& 2X_{\a(2s-2)} \big|~ , \\
 y_{\a(2s-1)} &:=& -\frac{\ri}{2}  D_{(\a_1 } X_{\a_2 \dots \a_{2s-1} )}\big|~, 
 \qquad 
 y_{\a(2s-3)} := -\ri  D^\b X_{\b \a (2s-3 )}\big| ~, \\
F_{\a(2s-2)}&:=& \frac{\ri}{4} X_{\a(2s-2)}\big|~.
\eea 
\end{subequations}
The gauge transformation laws of $ y_{\a(2s-2)}$, 
$ y_{\a(2s-1)}$ and $y_{\a(2s-3)} $ are as follows:
\begin{subequations}
\bea
\d_\z y_{\a (2s-2) } &=& \frac{2s}{2s+1} \pa^{\b\g } \z_{\b \g \a (2s-2) }~,  \label{7.9a}  \\
\d_\x y_{\a(2s-1)} &=&\frac{1}{2s+1}  \pa^\b{}_{(\a_1} \x_{\a_2 \dots \a_{2s-1}) \b}~,
 \label{7.9b}  \\
\d_\x y_{\a(2s-3)} &=& \pa^{\b \g} \x_{\b\g  \a (2s-3)}~.  \label{7.9c} 
\eea 
\end{subequations}
In principle, we do not need to derive the gauge transformation of $F_{\a(2s-2)}$ 
since this field turns out to be auxiliary.

The bosonic transformation laws \eqref{7.7a} and \eqref{7.9a} correspond to the 
massless spin-$(s+1)$ action $S_{\rm{F}}^{(2s+2)}$ 
defined by eq. \eqref{3.29}. The fermionic  transformation laws \eqref{7.7b},
 \eqref{7.9b}  and \eqref{7.9c} correspond to the 
massless spin-$(s+\hf)$ action $S_{\rm{FF}}^{(2s+1)}$ 
defined by eq. \eqref{3.6}.

The component action follows from \eqref{7.2} by making use of
the reduction rule \eqref{6.1}.
Direct calculations lead to the following bosonic Lagrangian:
\bea
2(-2)^{s+1} L_{\rm bos} &=& 
h^{\a(2s+2)}\Box h_{\a(2s+2)}
-\hf (s+1)\partial_{\g(2)}h^{\g(2)\a(2s)}\partial^{\b(2)}h_{\a(2s)\b(2)}\notag\\
&& -\hf{(2s-1)} y^{\a(2s-2)}\partial^{\b(2)}\partial^{\g(2)}h_{\a(2s-2)\b(2)\g(2)}
-\frac{(s+1)(2s-1)}{2s}
y^{\a(2s-4)}\Box y_{\a(2s-4)}  \non \\
&& -4 s(2s-1) \Big[ (s+1) F^{\a(2s-2)} F_{\a(2s-2)}
-\frac{s-1}{2s} F^{\a(2s-2)} \pa^\b{}_{(\a_1} y_{\a_2 \dots \a_{2s-2})\b} \Big]~.
\eea
Eliminating the auxiliary field $F_{\a(2s-2)} $  leads to 
\bea
2(-2)^{s+1} L_{\rm bos} &=& 
h^{\a(2s+2)}\Box h_{\a(2s+2)}
-\hf (s+1)\partial_{\g(2)}h^{\g(2)\a(2s)}\partial^{\b(2)}h_{\a(2s)\b(2)}\notag\\
&&-\hf{(2s-1)}\Big[ y^{\a(2s-2)}\partial^{\b(2)}\partial^{\g(2)}h_{\a(2s-2)\b(2)\g(2)}
+ \frac{2}{s+1}y^{\a(2s-2)}\Box y_{\a(2s-2)}\notag\\
&&+\frac{(s-1)(2s-3)}
{4(s+1)}\partial_{\g(2)}y^{\g(2)\a(2s-4)}\partial^{\b(2)}y_{\b(2)\a(2s-4)}\Big]
~.
\eea
This Lagrangian corresponds to the massless spin-$(s+1)$ action 
$S_{\rm{F}}^{(2s+2)}$ obtained from  \eqref{3.29} by the replacement $s \to s+1$.
The fermionic sector of the component action proves to coincide with 
the massless spin-$(s+\hf)$ action, 
$ S_{\rm{FF}}^{(2s+1)} [{h}_{(2s+1)},{y}_{(2s-1)},{ y}_{(2s-3)}]$.


\subsection{Massive  half-integer superspin action}

We now have all of the ingredients at our disposal to read off the component form of 
the massive half-integer superspin action that is obtained 
from \eqref{5.39b} in the flat-superspace limit,
\bea
{\mathbb S}_{\rm massive}^{(2s+1)}
&=&\phantom{+}
\lambda {\mathbb S}_{\rm{SCS}}^{(2s+1)}[{ H}_{(2s+1)}]
+\mu^{2s-1}{\mathbb S}_{\rm{SO}}^{(2s+1)}
[{ H}_{(2s+1)},{ X}_{(2s-2)}] \non \\
& \approx & \phantom{+}
\l S^{(2s+2)}_{\rm CS}[h_{(2s+2)}]  
+\mu^{2s-1} S_{\rm{F}}^{(2s+2)}
[{h}_{(2s+2)},{y}_{(2s-2)}]
\non \\
&&+ \l S^{(2s+1)}_{\rm CS}[h_{(2s+1)}] + 
\mu^{2s-1} S_{\rm{FF}}^{(2s+1)}
[{h}_{(2s+1)},{y}_{(2s-1)},{ y}_{(2s-3)}]~.
\label{6.34}
\eea
Here the symbol `$\approx$' indicates that the auxiliary field has been eliminated.

The explicit  structure of the component action  \eqref{6.34}
clearly demonstrates that the model 
\bea
{\mathbb S}_{\rm massive}^{(2s+1)}
&=&\lambda {\mathbb S}_{\rm{SCS}}^{(2s+1)}[{H}_{(2s+1)}]
+\mu^{2s-1}{\mathbb S}_{\rm{SO}}^{(2s+1)}
[{H}_{(2s+1)},{X}_{(2s-2)}]
\eea
describes a single massive supermultiplet subject to the equations
\eqref{214a} and \eqref{214b}  with $n=2s+1$ on the mass shell.
The superfield proof was provided in \cite{KT}.


\section{Concluding comments}

All massive higher-spin theories in Minkowski space, 
which have been presented in this 
paper, were extracted from off-shell supersymmetric field theories. As shown in section 
6, all the theories studied in section 4 are contained at the component level 
in the  $\cN=1$ supersymmetric massive higher-spin theories 
proposed in \cite{KT}. The latter models were obtained 
from the   $\cN=2$ supersymmetric massive higher-spin theories 
of \cite{KO} by carrying out the $\cN=2 \to \cN=1$ superspace reduction. 
Furthermore, the off-shell structure of the massless 3D $\cN=2$ supersymmetric higher-spin actions of \cite{KO}, which constitute one of the two sectors of 
the $\cN=2$ massive actions, were designed following the pattern of 
the gauge off-shell formulations for massless 4D $\cN=1$ 
higher-spin supermultiplets  developed in the early 1990s \cite{KSP,KS}.

Our  supersymmetric massive higher-spin theories, which are formulated in 
AdS${}^{3|2}$ superspace and are described by the actions 
\eqref{5.39a} and \eqref{5.39b},
contain two different models for a massive integer-spin field in AdS
at the component level.
One of them is the gauge-invariant model \eqref{3.44b}.
The second model is described by the action 
\bea
\widetilde{S}_{\rm massive}^{(2s)}
&=&\lambda S_{\rm{CS}}^{(2s)}[{\mathfrak h}_{(2s)}]
+\hf \Big( -\hf\Big)^s \mu^{2s-1}
\int \rd^3 x \, e\, {\mathfrak h}^{\a (2s)} {\mathfrak h}_{\a (2s)} ~,
\label{7.1}
\eea
which does not possess gauge invariance and which is the AdS uplift of the model \eqref{6.19}. The  action \eqref{7.1} leads to the equation of motion
\bea
\l {\mathfrak C}_{\a(2s)} + \mu^{2s-1}  {\mathfrak h}_{\a (2s)}=0
\quad \implies \quad \nabla^{\b\g} {\mathfrak h}_{\b\g\a (2s-2)}=0
~.
\eea
The action \eqref{7.1} can be turned into a gauge-invariant one by 
making use of the St\"uckelberg trick.
An interesting feature of the model  \eqref{7.1} is that it is well-defined
in an arbitrary conformally flat space.

The models \eqref{3.44a} and \eqref{7.1} are higher-spin analogues
of the two well-known equivalent models for a massive vector field
(see \cite{TPvN,DJ} and references therein)
with Lagrangians
\begin{subequations}
\bea
\cL_{\rm T} &=& -\frac 14 F^{ab}F_{ab} +\frac{m}{4} \ve^{abc} V_a F_{bc}~, 
\qquad F_{ab}=\pa_a V_b - \pa_b V_a~,
\\
\cL_{\rm SD} &=& \hf f^a f_a -\frac{1}{2m} \ve^{abc} f_a \pa_b f_c~.
\eea
\end{subequations}

New duality transformations were introduced in \cite{K16} for theories 
formulated in terms of the linearised higher-spin Cotton tensors
$C_{\a(n)}$ and their $\cN=1$ supersymmetric counterparts $W_{\a(n)}$. 
These duality transformations can readily be generalised to arbitrary 
conformally flat backgrounds, with $C_{\a(n)}$  and  $W_{\a(n)}$ 
replaced with ${\mathfrak C}_{\a(n)}$ and  ${\mathfrak W}_{\a(n)}$,
respectively.

In the present paper, we have been unable to obtain closed-form 
expressions for ${\mathfrak C}_{\a(n)}$ and  ${\mathfrak W}_{\a(n)}$ 
in terms of the covariant derivatives of AdS (super)space
for arbitrary $n$. These are interesting open problems.

The field strengths ${\mathfrak C}_{\a(n)}$ and  ${\mathfrak W}_{\a(n)}$
are the higher-spin extensions of the linearised 
Cotton and super-Cotton tensors, respectively. 
The actions \eqref{2.29} and \eqref{5.21} are the higher-spin 
extensions of the linearised actions for conformal gravity and 
supergravity, respectively. An intriguing question is: Do nonlinear 
higher-spin extensions exist? Within the approach initiated in
\cite{Nilsson1,Nilsson2}, Linander and Nilsson \cite{LN}
constructed the full nonlinear spin-3 Cotton equation coupled to 
spin-2. They made use of the frame field description 
and the Chern-Simons formulation for  3D (super)conformal 
field theory due to Fradkin and Linetsky \cite{FL}. 
The construction of the nonlinear spin-3 Cotton tensor \cite{LN} requires an elimination 
of certain auxiliary fields, a procedure that becomes extremely difficult
for $s>3$.  However, so far this is unexplored territory.
There exist nonlinear formulations for the massless spin-3 theory
\cite{HR,CFPT}, and the generalisation from $s=3$ to $s>3$ 
is shown in \cite{CFPT} to be trivial within the formulation developed. 
These results indicate that it is possible to construct a nonlinear 
topologically massive higher-spin field theory. 
The fundamental results by Prokushkin and Vasiliev
\cite{PV1,PV2} should be essential of course.
Any attempt to construct a supersymmetric interacting higher-spin theory 
should inevitably be an extension of the conformal superspace approach 
\cite{BKNT-M1,BKNT-M2}.

It should be pointed out that the problem of constructing topologically massive 
higher-spin theories was considered in  \cite{Chen1,Chen2}. However, 
the nonlinear action proposed possesses only a restricted gauge freedom 
in the presence of the Lagrange multiplier $\b$ that enforces the torsion-free conditions
on the spin connections. Alternative approaches are worth pursuing. 

So far we have discussed $\cN=1$ topologically massive supergravity 
and its higher spin extensions.
The off-shell formulations for $\cN$-extended topologically massive supergravity theories were presented 
in \cite{DKSS2,KLRST-M} for $\cN=2$,  in \cite{KN14} for $\cN=3$,
and in \cite{KN14,KNS} for the $\cN=4$ case.  
In all of these theories, the action functional is a sum of two terms, 
one of which is the action for pure $\cN$-extended supergravity 
(Poincar\'e or anti-de Sitter)
and the other is the action for $\cN$-extended conformal supergravity.
The off-shell actions for $\cN$-extended supergravity theories 
in three dimensions were given
in \cite{GGRS} for $\cN=1$, \cite{KLT-M11,KT-M11} for $\cN=2$ and
 \cite{KLT-M11} for the cases $\cN=3, ~4$.
The off-shell actions for $\cN$-extended conformal supergravity 
were given in \cite{vN} for $\cN=1$, \cite{RvanN86} for $\cN=2$,  
  \cite{BKNT-M2} for $ \cN =3,~4,~ 5$, and in \cite{NT,KNT-M13}
for  the $\cN=6$ case.
 Refs. \cite{BKNT-M2,KNT-M13} made use of
the off-shell formulation for $\cN$-extended conformal supergravity 
proposed in \cite{BKNT-M1}. The on-shell formulation for $\cN$-extended
conformal supergravity with $\cN>2$ was given in \cite{LR89}.
On-shell approaches to  $\cN$-extended topologically massive supergravity theories 
with $4 \leq \cN \leq 8$ were presented in  
  \cite{Chu:2009gi,Gran:2012mg,Nilsson:2013fya,LS1,LS2}. 
It would be interesting to formulate topologically massive higher spin supermultiplets
for $\cN>2$.
\\

\noindent
{\bf Note added in proof:}\\
The  equations  \eqref{2.33ab} for massive fields in AdS${}_3$
may be realised as equations 
of motion in the following model 
\bea
S_{\rm{massive}}^{(n)} [ {\mathfrak h}_{(n)}] 
=\frac{\text{i}^n}{2^{\left \lfloor{n/2}\right \rfloor +1}}
\frac{\l}{\m}  
\int \rd^3 x \, e\, {\mathfrak C}^{\a(n) } ( {\mathfrak h})
\Big\{ \m \d^\b{}_{\a_1} +\nabla^\b{}_{\a_1} \Big\}  {\mathfrak h}_{\a_2 \dots \a_n \b} 
~, 
\non
\eea
which is invariant under the gauge transformations \eqref{3.155} in AdS${}_3$.
It is ${\mathfrak C}_{\a(n) } ( {\mathfrak h})$ which plays the  role of $\f_{\a(n)}$. 
The  equations  \eqref{2.12ab} for massive superfields in AdS${}^{3|2}$
 may be realised as equations 
of motion in the following model 
\bea
{\mathbb S}_{\rm{massive}}^{(n)} [ {\mathfrak H}_{(n)}] 
= - \frac{\ri^n}{2^{\left \lfloor{n/2}\right \rfloor +1}}
\frac{\l}{\m}   \int \rd^{3|2}z \, E\,{\mathfrak W}^{\a(n) }( {\mathfrak H})
\Big\{  \m + \frac{\ri}{2} \cD^2 \Big\}{\mathfrak H}_{\a(n)} 
 ~,
\non
\eea
which is invariant under the gauge transformations \eqref{5.122}
in AdS${}^{3|2}$.
It is ${\mathfrak W}_{\a(n) } ( {\mathfrak H})$ which plays the  role of $T_{\a(n)}$. 
These models, which become (super)conformal in the $\m \to\infty$ limit,
 may be viewed as generalisations of the flat-space bosonic
constructions of \cite{BHT,BKRTY}.
\\


\noindent
{\bf Acknowledgements:}\\
SMK is grateful to Mirian Tsulaia  for collaboration at the initial stages of the project,
as well as for pointing out important references.
The work of SMK is supported in part by the Australian 
Research Council, project No. DP160103633.
The work of MP is supported by the Hackett Postgraduate Scholarship UWA ,
under the Australian Government Research Training Program.

\appendix

\section{Notation and conventions}
We follow the notation and conventions adopted in
\cite{KLT-M11}. In particular, the Minkowski metric is
$\eta_{ab}=\mbox{diag}(-1,1,1)$.
The spinor indices are  raised and lowered using
the $\rm SL(2,{\mathbb R})$ invariant tensors
\bea
\ve_{\a\b}=\left(\begin{array}{cc}0~&-1\\1~&0\end{array}\right)~,\qquad
\ve^{\a\b}=\left(\begin{array}{cc}0~&1\\-1~&0\end{array}\right)~,\qquad
\ve^{\a\g}\ve_{\g\b}=\d^\a_\b
\eea
by the standard rule:
\bea
\psi^{\a}=\ve^{\a\b}\psi_\b~, \qquad \psi_{\a}=\ve_{\a\b}\psi^\b~.
\label{A2}
\eea

We make use of real gamma-matrices,  $\g_a := \big( (\g_a)_\a{}^\b \big)$, 
which obey the algebra
\be
\gamma_a \gamma_b=\eta_{ab}{\mathbbm 1} + \varepsilon_{abc}
\gamma^c~,
\label{A3}
\ee
where the Levi-Civita tensor is normalised as
$\varepsilon^{012}=-\varepsilon_{012}=1$. The completeness
relation for the gamma-matrices reads
\be
(\gamma^a)_{\alpha\beta}(\gamma_a)^{\rho\sigma}
=-(\delta_\alpha^\rho\delta_\beta^\sigma
+\delta_\alpha^\sigma\delta_\beta^\rho)~.
\label{A4}
\ee
Here the symmetric matrices 
$(\gamma_a)^{\alpha\beta}$ and $(\gamma_a)_{\alpha\beta}$
are obtained from $\g_a=(\g_a)_\a{}^{\b}$ by the rules (\ref{A2}).
Some useful relations involving $\g$-matrices are 
\bsubeq
\bea
\ve_{abc}(\g^b)_{\a\b}(\g^c)_{\g\d}&=&
\ve_{\g(\a}(\g_a)_{\b)\d}
+\ve_{\d(\a}(\g_a)_{\b)\g}
~,
\\
\tr[\g_a\g_b\g_{c}\g_d]&=&
2\eta_{ab}\eta_{cd}
-2\eta_{ac}\eta_{db}
+2\eta_{ad}\eta_{bc}
~.
\eea
\esubeq

Given a three-vector $x_a$,
it  can be equivalently described by a symmetric second-rank spinor $x_{\a\b}$
defined as
\bea
x_{\a\b}:=(\g^a)_{\a\b}x_a=x_{\b\a}~,\qquad
x_a=-\hf(\g_a)^{\a\b}x_{\a\b}~.
\eea
In the 3D case,  an
antisymmetric tensor $F_{ab}=-F_{ba}$ is Hodge-dual to a three-vector $F_a$, 
specifically
\bea
F_a=\hf\ve_{abc}F^{bc}~,\qquad
F_{ab}=-\ve_{abc}F^c~.
\label{hodge-1}
\eea
Then, the symmetric spinor $F_{\a\b} =F_{\b\a}$, which is associated with $F_a$, can 
equivalently be defined in terms of  $F_{ab}$: 
\bea
F_{\a\b}:=(\g^a)_{\a\b}F_a=\hf(\g^a)_{\a\b}\ve_{abc}F^{bc}
~.
\label{hodge-2}
\eea
These three algebraic objects, $F_a$, $F_{ab}$ and $F_{\a \b}$, 
are in one-to-one correspondence to each other, 
$F_a \leftrightarrow F_{ab} \leftrightarrow F_{\a\b}$.
The corresponding inner products are related to each other as follows:
\bea
-F^aG_a=
\hf F^{ab}G_{ab}=\hf F^{\a\b}G_{\a\b}
~.
\eea

The Lorentz generators with two vector indices ($M_{ab} =-M_{ba}$),  one vector index ($M_a$)
and two spinor indices ($M_{\a\b} =M_{\b\a}$) are related to each other by the rules:
$M_a=\hf \ve_{abc}M^{bc}$ and $M_{\a\b}=(\g^a)_{\a\b}M_a$.
These generators 
act on a vector $V_c$ 
and a spinor $\J_\g$ 
as follows:
\bea
M_{ab}V_c=2\eta_{c[a}V_{b]}~, ~~~~~~
M_{\a\b}\J_{\g}
=\ve_{\g(\a}\J_{\b)}~.
\label{generators}
\eea


\section{First-order higher-spin model}

In this appendix we review the first-order higher-spin model in Minkowski space 
used by Tyutin and Vasiliev
\cite{TV} in their formulation for massive higher-spin fields. 
It is realised in terms of a 
reducible  field ${\bm h}_{b, \a_1 \dots \a_{n-2}} = {\bm h}_{b, (\a_1 \dots \a_{n-2})} $
which is defined modulo 
gauge transformations of the form 
\bea
\d {\bm h}_{b, \a_1 \dots \a_{n-2}}  = \pa_b \x_{ \a_1 \dots \a_{n-2}} ~, 
\qquad \x_{ \a_1 \dots \a_{n-2}} = \x_{ (\a_1 \dots \a_{n-2} )}~.
\label{A.1}
\eea 
The structure of this transformation implies that the following  action 
\bea
S_{\rm{FF}}^{(n)}= -\frac{\text{i}^n}{2^{\lfloor n/2 \rfloor }} \int \rd^3 x \, \ve^{bcd} {\bm h}_{b,}{}^{\a_1 \dots \a_{n-2}} \pa_c 
{\bm h}_{d, \a_1 \dots \a_{n-2}}
\label{TVaction}
\eea
is gauge invariant.

The field ${\bm h}_{\b \g, \a_1 \dots \a_{n-2}} := (\g^b)_{\b\g} {\bm h}_{b , \a_1 \dots \a_{n-2}}$  
contains three irreducible ${\rm SL}(2,{\mathbb R})$ fields that we define as follows:
\begin{subequations}
\bea
h_{\a_1 \dots \a_{n} } &:=& {\bm h}_{(\a_1 \a_2  , \a_3 \dots \a_{n})}~, \\
y_{\a_1 \dots \a_{n-2} } &:=& 
\frac{1}{n} {\bm h}^\b{}_{(\a_1 , \a_2 \dots \a_{n-2}) \b}
~,  \\
y_{\a_1 \dots \a_{n-4} } &:=& {\bm h}^{\b \g ,}{}_{\b \g \a_1 \dots \a_{n-4}}~.
\eea
\end{subequations}
In accordance with \eqref{A.1},
the gauge transformation laws of these fields are 
\begin{subequations}
\bea
\d h_{\a_1 \dots \a_{n}} &=& \pa_{(\a_1 \a_2} \x_{\a_3 \dots \a_{n})} ~, \\
\d y_{\a_1 \dots \a_{n-2}} &=&\frac{1}{n} \pa^\b{}_{(\a_1 } \x_{\a_2 \dots \a_{n-2})\b}~, \\
\d y_{\a_1 \dots \a_{n-4}} &=& \pa^{\b  \g} \x_{\b \g \a_1 \dots \a_{n-4}}~.
\eea
\end{subequations}
The action \eqref{TVaction} turns into 
\bea
S_{\rm{FF}}^{(n)}&=& \frac{\text{i}^n}{2^{\lfloor n/2 \rfloor +1}}
 \int \rd^3 x \, \Big\{ 
h^{\b \a_1 \dots \a_{n-1} }\pa_\b{}^\g h _{\g \a_1 \dots \a_{n-1}} 
+2(n-2)
y^{\a_1 \dots \a_{n-2}} \pa^{\b \g} h_{\b \g \a_1 \dots \a_{n-2}} \non \\
&& + 4(n-2)y^{\b \a_1 \dots \a_{n-3} }
\pa_\b{}^\g y _{\g \a_1 \dots \a_{n-3}} 
+2\frac{n(n-3)}{n-1} y^{\a_1 \dots \a_{n-4}} \pa^{\b \g} y_{\b \g \a_1 \dots \a_{n-4}} \non \\
&& - \frac{(n-3)(n-4)}{(n-1)(n-2)}  y^{\b \a_1 \dots \a_{n-3} }
\pa_\b{}^\g y _{\g \a_1 \dots \a_{n-3}} \Big\}~.
\label{B.5}
\eea
This is the flat-space limit of the first-order action \eqref{33.41}.
When $n$ is odd, $n=2s+1$, 
the functional $S_{\rm{FF}}^{(2s+1)}$ coincides with 
 plain $\rm 4D \to 3D $ dimensional reduction 
of the Fang-Fronsdal action \cite{FF}.


\section{Higher-spin Cotton tensor as a descendent of gauge-invariant field strengths}

The Cotton tensor is defined in terms of the Ricci tensor according to \eqref{Cotton33}. The latter determines the equations of motion corresponding to the Einstein-Hilbert action. In this appendix we show that analogous properties hold 
for the linearised higher-spin Cotton tensor defined by eq. \eqref{2.31}.


\subsection{The first-order case}

We begin by demonstrating that 
the higher-spin Cotton tensor  \eqref{2.31}
is a descendant of gauge-invariant field strengths which determine 
the equations of motion in the first-order model \eqref{B.5}.
Associated with the dynamical variables $h_{\a(n)},y_{\a(n-2)}$ and $y_{\a(n-4)}$
are the following gauge-invariant field strengths:
\bsubeq
\bea
F_{\a(n)}&:=&\partial_{(\a_1}{}^{\b}h_{\a_2\dots\a_{n})\b}-(n-2)\partial_{(\a_1\a_2}y_{\a_3\dots\a_{n})}~, \label{C.1a}\\
G_{\a(n-2)}&:=&\partial^{\b(2)}h_{\a(n-2)\b(2)}+4\partial_{(\a_1}{}^{\b}y_{\a_2\dots\a_{n-2})\b}-\frac{n(n-3)}{(n-1)(n-2)}\partial_{(\a_1\a_2}y_{\a_3\dots \a_{n-2})}~,~~~~~~ \label{C.1b}\\
H_{\a(n-4)}&:=&(n-2)\partial^{\b(2)}y_{\a(n-4)\b(2)}-\frac{n-4}{n}\partial_{(\a_1}{}^{\b}y_{\a_2\dots\a_{n-4})\b}~. \label{C.1c}
\eea
\esubeq
The equations of motion corresponding to \eqref{B.5} are the conditions 
that these field strengths vanish. Furthermore, the gauge symmetry implies that  $F_{\a(n)},~G_{\a(n-2)}$ and $ H_{\a(n-4)}$ are related to each other via the  
Noether identity
\bea
0=\partial^{\b(2)}F_{\a(n-2)\b(2)}-\frac{n-2}{n}\partial_{(\a_1}{}^{\b}G_{\a_2\dots\a_{n-2})\b}+\frac{n(n-3)}{(n-1)(n-2)}\partial_{(\a_1\a_2}H_{\a_3\dots\a_{n-2})}~. \label{C.2}
\eea

We claim that the Cotton tensor $C_{\a(n)} (h)$
may be expressed as 
$C_{\alpha(n)}=(\mathcal{A}_1F)_{\a(n)}+(\mathcal{A}_2G)_{\a(n)}+(\mathcal{A}_3H)_{\a(n)}$, for some linear differential operators $\mathcal{A}_i$ of order $n-2$. A suitable ansatz for such an expression is
\begin{align}
C_{\a(n)}&=\sum_{j=0}^{\lfloor\frac{n}{2}\rfloor-1}a_j\square^j\partial_{(\a_1}{}^{\b_1}\cdots\partial_{\a_{n-2j-2}}{}^{\b_{n-2j-2}}F_{\a_{n-2j-1}\dots\a_{n})\b_1\dots\b_{n-2j-2}} \notag\\
&+\sum_{k=0}^{\lceil \frac{n}{2} \rceil-2}b_k\square^k\partial_{(\a_1}{}^{\b_1}\cdots\partial_{\a_{n-2k-3}}{}^{\b_{n-2k-3}}\partial_{\a_{n-2k-2}\a_{n-2k-1}}G_{\a_{n-2k}\dots\a_{n})\b_1\dots\b_{n-2k-3}}\label{C.3}\\
&+\sum_{l=0}^{\lfloor\frac{n}{2}\rfloor-2}c_l\square^l\partial_{(\a_1}{}^{\b_1}\cdots\partial_{\a_{n-2l-4}}{}^{\b_{n-2l-4}}\partial_{\a_{n-2l-3}\a_{n-2l-2}}
\notag\\
&
\qquad \qquad 
\times 
\partial_{\a_{n-2l-1}\a_{n-2l}}
H_{\a_{n-2l+1}\dots\a_{n})\b_1\dots\b_{n-2l-4}} \notag
\end{align}
for some coefficients $a_j,b_k$ and $c_l$. It may be shown that the values of these coefficients are not unique and that there are $\lfloor\frac{n}{2}\rfloor-1$ free parameters.  For example, when $n=5$ one may show that the general solution is 
\bea
\begin{pmatrix}
a_0\\a_1\\b_0\\b_1\\c_0
\end{pmatrix} =\begin{pmatrix}
\frac{1}{2}+\frac{18}{5}c_0\\\frac{1}{2}-\frac{18}{5}c_0\\\frac{9}{80}-\frac{36}{25}c_0\\\frac{3}{80}-\frac{18}{25}c_0\\c_0
\end{pmatrix}~. \notag
\eea
We may use this freedom to completely eliminate the $\lfloor\frac{n}{2}\rfloor-1$ coefficients $c_l$ so that only the 
field strengths $F_{\a(n)}$ and $G_{\a(n-2)}$ appear in \eqref{C.3}. This fixes the solution uniquely to
\bsubeq
\bea
a_j&=&\frac{1}{2^{n-2}}\frac{(n-1)}{(2j+1)}\binom{n-2}{2j} \qquad \qquad ~~~~ \text{for}~~ 0\leq j \leq \bigg\lfloor\frac{n}{2}\bigg\rfloor-1~, \label{C.4a}\\[2 ex]
b_k&=&\frac{1}{2^{n-1}}\frac{(n-2)^2}{n(2k+1)}\binom{n-3}{2k} \qquad\qquad ~~\text{for}~~0\leq k \leq \bigg\lceil \frac{n}{2} \bigg\rceil-2~, \label{C.4b} \\[2ex]
c_l&=&0 \phantom{BLANK SPACEEEEEEEEE}~ \text{for}~~ 0\leq l \leq \bigg\lfloor\frac{n}{2}\bigg\rfloor-2~.
\eea
\esubeq
The fact that there are $\lfloor\frac{n}{2}\rfloor-1$ free parameters may be understood as a consequence of the Noether identity \eqref{C.2}. To see this, observe that, in principle, we may use \eqref{C.2} to replace all occurrences of $H_{\a(n-4)}$ with $F_{\a(n)}$ and $G_{\a(n-2)}$ in the ansatz \eqref{C.3}. There will then be only two sets of independent coefficients, say $\tilde{a}_j$ and $\tilde{b}_k$, 
whose unique  values coincide with those of \eqref{C.4a} and \eqref{C.4b}.  


\subsection{The second-order case}

We now consider the flat-space version of 
the second-order model \eqref{33.43}.
It is described by the real fields $h_{\a(n)}$ and $h_{\a(n-4)}$. 
Associated with these two fields are the following gauge-invariant field strengths:
\bsubeq 
 \begin{align}
 F_{\a(n)}&=\square h_{\a(n)}+\frac{n}{4}\partial^{\b(2)}\partial_{(\a_1\a_2}h_{\a_3\dots\a_{n})\b(2)}-\frac{n-3}{4}\partial_{(\a_1\a_2}\partial_{\a_3\a_4}y_{\a_5\dots\a_{n})}~,\label{C.5a}\\
 G_{\a(n-4)}&= \partial^{\b(2)}\partial^{\b(2)}h_{\a(n-4)\b(4)}+\frac{8}{n}\square y_{\a(n-4)}-\frac{(n-4)(n-5)}{n(n-2)}\partial^{\b(2)}\partial_{(\a_1\a_2}y_{\a_3\dots\a_{n-4})\b(2)}~.~~~~~~~~~~~\label{C.5b}
 \end{align}
 \esubeq
 The equations of motion for the model
   are $F_{\a(n)}=0$ and $G_{\a(n-4)}=0$. The two field strengths are related by the Noether identity
 \bea
 \partial^{\b(2)}F_{\a(n-2)\b(2)}=\frac{(n-3)(n-2)}{4(n-1)}\partial_{(\a_1\a_2}G_{\a_3\dots\a_{n-2})}~.
 \eea
 We claim that the Cotton tensor $C_{\a(n)} (h)$
 may be written as 
 $C_{\alpha(n)}=(\mathcal{A}_1F)_{\a(n)}+(\mathcal{A}_2G)_{\a(n)}$ where the $\mathcal{A}_i$ are linear differential operators of order $n-3$. A suitable ansatz for such an expression is
\begin{align}
C_{\a(n)}=&\sum_{j=0}^{\lceil \frac{n}{2} \rceil-2}a_j\square^j\partial_{(\a_1}{}^{\b_1}\cdots\partial_{\a_{n-2j-3}}{}^{\b_{n-2j-3}}F_{\a_{n-2j-2}\dots\a_{n})\b_1\dots\b_{n-2j-3}} \label{C.7}\\
&+\sum_{k=0}^{\lceil \frac{n}{2} \rceil-3}b_k\square^k\partial_{(\a_1}{}^{\b_1}\cdots\partial_{\a_{n-2k-5}}{}^{\b_{n-2k-5}}
\non \\
& \qquad 
\times
\partial_{\a_{n-2k-4}\a_{n-2k-3}}\partial_{\a_{n-2k-2}\a_{n-2k-1}}
G_{\a_{n-2k}\dots\a_{n})\b_1\dots\b_{n-2k-5}}~,
\notag~
\end{align}
for some coefficients $a_j$ and $b_k$. It may be shown that the choice of these coefficients is not unique, and that there are $\lceil \frac{n}{2} \rceil-2$ free parameters. For example, when $n=6$ one may show that the general solution is 
\bea
\begin{pmatrix}
a_0\\a_1\\b_0
\end{pmatrix} =
\begin{pmatrix}
\frac{5}{8}-\frac{10}{3}b_0 \\ \frac{3}{8}+\frac{10}{3}b_0\\b_0
\end{pmatrix}~. \notag
\eea
We can use this freedom to completely eliminate the $\lceil \frac{n}{2} \rceil-2$ coefficients $b_k$ so that only the top field strength, $F_{\a(n)}$, appears in \eqref{C.7}. This gives the unique solution
\bsubeq
\bea
a_j&=&(j+1)\frac{\binom{n-3}{2j}}{\binom{2j+3}{3}}\frac{n(n-1)}{3\cdot 2^{n-2}}~\phantom{BLANK SPA} \text{for}~~0\leq j\leq \bigg\lceil \frac{n}{2} \bigg\rceil-2~,\\
b_k&=&0 \phantom{BLAN SPACEEEEEEEEE}~~~~~~~ \text{for}~~ 0\leq k\leq \bigg\lceil \frac{n}{2} \bigg\rceil-3~.
\eea
\esubeq


\begin{footnotesize}

\end{footnotesize}

\end{document}